%% file: main.tex
\documentclass[twocolumn,prd, aps, notitlepage, nofootinbib, showpacs, superscriptaddress,longbibliography]{revtex4-1}
\usepackage[utf8]{inputenc}
\usepackage{graphicx,mathtools,xcolor,latexsym,rotating,soul}
\usepackage[T1]{fontenc}
\usepackage{tikz}
\usepackage{comment}
\usepackage{amsmath, amssymb, amsthm, amsfonts,breqn, bbm}
\usepackage{multirow}
\usetikzlibrary{decorations.pathmorphing}
\usepackage[colorlinks=true,citecolor=green,urlcolor=blue,linkcolor=blue]{hyperref}
\usepackage{pifont}
\usepackage{gensymb}
\usepackage{enumitem}
\usepackage{mathrsfs}
\usepackage{array}

\newtheoremstyle{named}{}{}{\itshape}{}{\bfseries}{.}{.5em}{\thmnote{#3 }#1}
\theoremstyle{named}

\newcolumntype{C}[1]{>{\centering\arraybackslash}p{#1}}
\newcommand*{\rom}[1]{\expandafter\@slowromancap\romannumeral ##x1@}
\newcommand{\thetanpe}{\theta_{\mathrm{NPE}}}
\newcommand{\thetaquark}{\theta_{\mathrm{quark}}}

\newcommand\T{\rule[3.5 ex] {0pt}{0 ex}}       
\newcommand\B{\rule[-2.5 ex]{0pt}{0pt}}
\newcommand{\stf}[1]{\langle #1 \rangle}

\newcommand{\Ga}{\Gamma}
\newcommand{\de}{\delta}

\newcommand{\om}{\omega}

\definecolor{MHcolor}{rgb}{0.65,0.10,0.10} 

\makeatletter
\let\cat@comma@active\@empty
\makeatother
\allowdisplaybreaks[4]
\begin{document}
\title{Dynamical tidal response of neutron stars as a probe of dense-matter properties}
\author{Abhishek Hegade K. R.}
\email{ah4278@princeton.edu}
\affiliation{Princeton Gravity Initiative, Princeton University, Princeton, NJ 08544, USA}
\affiliation{The Grainger College of Engineering, Illinois Center for Advanced Studies of the Universe, Department of Physics, University of Illinois at Urbana-Champaign, Urbana, IL 61801, USA}

\author{Yumu Yang}
\email{yumuy2@illinois.edu}
\affiliation{The Grainger College of Engineering, Illinois Center for Advanced Studies of the Universe, Department of Physics, University of Illinois at Urbana-Champaign, Urbana, IL 61801, USA}

\author{Mauricio Hippert}
\affiliation{Centro Brasileiro de Pesquisas Físicas, Rua Dr. Xavier Sigaud 150, 
Rio de Janeiro, RJ, 22290-180, Brazil}

\author{Jacquelyn Noronha-Hostler}
\affiliation{The Grainger College of Engineering, Illinois Center for Advanced Studies of the Universe, Department of Physics, University of Illinois at Urbana-Champaign, Urbana, IL 61801, USA}

\author{Jorge Noronha}
\affiliation{The Grainger College of Engineering, Illinois Center for Advanced Studies of the Universe, Department of Physics, University of Illinois at Urbana-Champaign, Urbana, IL 61801, USA}

\author{Nicol\'as Yunes}
\affiliation{The Grainger College of Engineering, Illinois Center for Advanced Studies of the Universe, Department of Physics, University of Illinois at Urbana-Champaign, Urbana, IL 61801, USA}

\begin{abstract}
Dynamical tidal deformations play a crucial role in the gravitational waves emitted by binary neutron star systems during their late inspiral.
In this work, we systematically explore how relativistic (dynamical and dissipative) tidal deformations depend on the internal structure of a neutron star using two analytic classes of equations of state. 
The first class is a nucleonic model that is parameterized by nuclear physics observables, such as the symmetry energy coefficients and saturation properties. 
The second class is a toy model of quark matter, the MIT bag model.
To model tidal dissipation, we self-consistently include contributions from weak-interaction-driven bulk-viscous effects while considering both the nucleonic and the quark-matter equations of state.
The dissipative tide is sensitive to frequency and temperature, but its magnitude, as predicted by weak-interaction-driven bulk-viscous effects, is too small (within the equation-of-state models studied here) to be detectable by current or future observations. 
However, we find that the (conservative) dynamical  tidal response function depends strongly on the slope of the symmetry energy and on higher-order coefficients of the symmetry energy. 
This indicates that the (conservative) dynamical tide is sensitive to higher-order coefficients, motivating a dedicated parameter-estimation study to assess how well they can be constrained.
\end{abstract}
\maketitle
\section{Introduction}
During the late inspiral of a binary neutron star coalescence, the stars are highly deformed by their companion's tidal field, and these deformations affect the gravitational waves emitted by the system. 
The rate of deformation depends sensitively on the properties of the dense, nuclear matter inside the stars \cite{Hinderer:2009ca,Nandi:2017rhy,Essick:2019ldf,Tan:2021ahl,Chatziioannou:2020pqz,Tan:2021nat,Legred:2025aar}.
Therefore, measuring gravitational waves emitted by inspiraling binary neutron stars and inferring their tidal deformabilities allows one to probe the supranuclear equation of state (EoS)~\cite{Flanagan:2007ix,LIGOScientific:2018cki,Chatziioannou_2020}.

During the early part of the late quasi-circular inspiral, tidal deformations are relatively weak, and the stars retain their equilibrium structure. 
A regime of the inspiral, therefore, exists in which one can assume that the tidal response function is approximately time independent.
Using this assumption, all the information about the internal properties of the neutron star can be encoded in a single coefficient $\Lambda$, called the (electric-type, quadrupolar) \textit{static tidal deformability}~\cite{Hinderer:2007mb,Binnington:2009bb,Damour:2009vw,Hinderer_2010}.
This tidal deformability coefficient affects the gravitational-wave phase at 5 post-Newtonian (PN) order\footnote{In the PN approximation, the Einstein equations are solved as an expansion in slow velocities and weak fields. A term of $N$ PN order is one that scales as $(v/c)^{2N}$ or $[GM/(c^2 d)]^N$ relative to the leading-order term, where $v$ is the orbital velocity of the binary, $M$ its total mass, and $d$ the orbital separation.}~\cite{Flanagan:2007ix}. Constraints from the GW170817 event \cite{LIGOScientific:2018hze} on $\Lambda$ have already provided some information on the cold, $\beta$-equilibrated EoS \cite{LIGOScientific:2018cki}, if one assumes the stars are not rapidly spinning.

\begin{figure}[htp]
    \centering
    \includegraphics[width=0.95\linewidth]{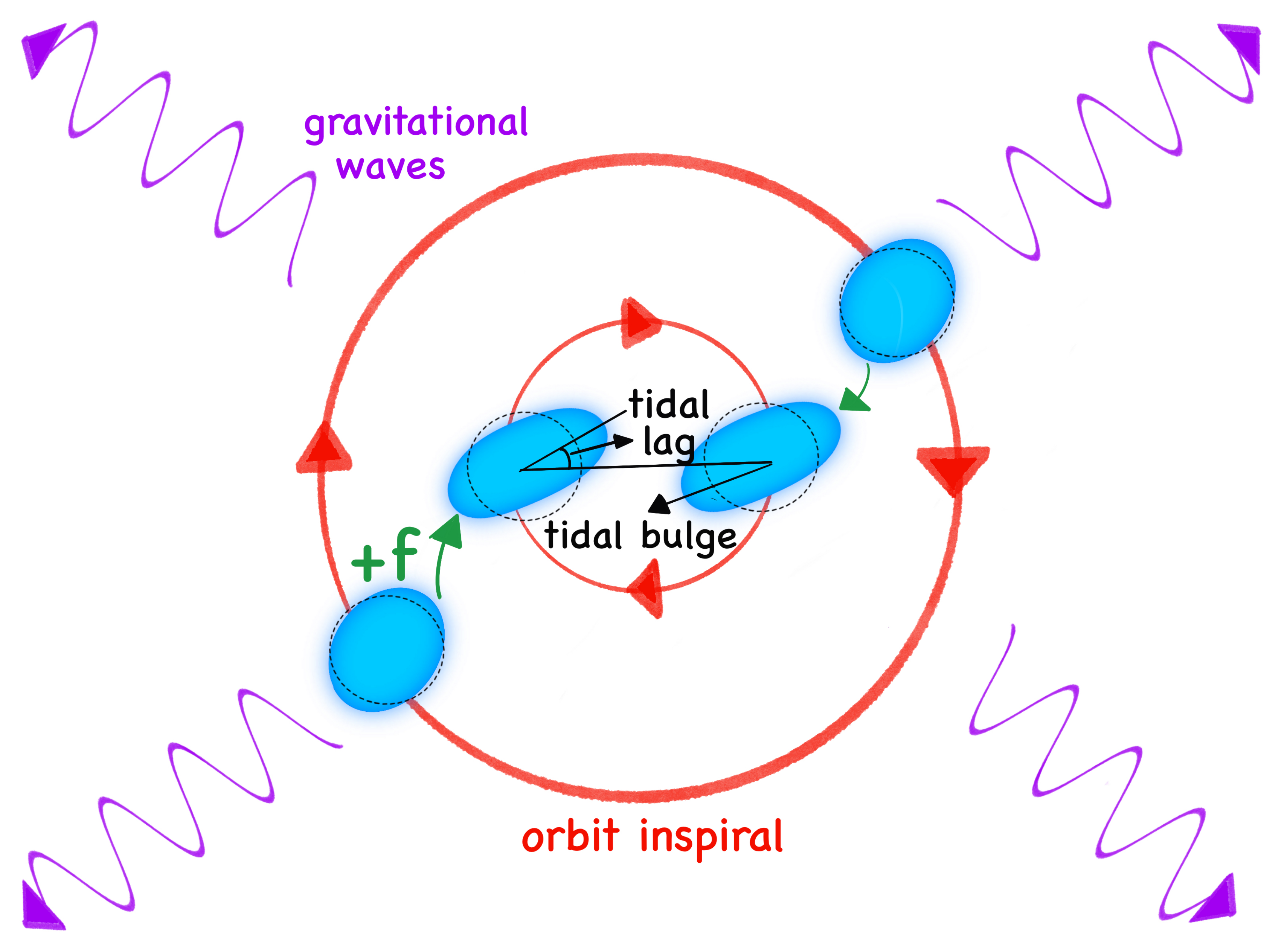}
    \caption{Cartoon of the tidal response of neutron stars in a binary system (not to scale). Qualitatively, the conservative response creates the tidal bulge and the dissipative response causes the tidal bulge to lag behind the gravitational field sourced by its companion. Both the conservative and the dissipative responses are enhanced as the orbital frequency $f$ increases due to gravitational wave emission.
    } 
    \label{fig:cartoon}
\end{figure}

The approximately quasi-static description of tidal dynamics breaks down during the very late inspiral (see Fig.~\ref{fig:cartoon}), and the tidal response function becomes complex and frequency dependent, with its real part called the \textit{conservative tidal response} and its imaginary part called the \textit{dissipative tidal response}.
The conservative tidal response reduces to the static tidal deformability (and thus, it is related to the static Love number $k_2$) in the zero frequency limit~\cite{Flanagan:2007ix,Steinhoff:2016rfi}.
The strongest frequency-dependent contribution to the conservative tidal response function occurs due to the $f$-mode of the neutron star, whose gravitational-wave frequency is typically of $\mathcal{O}(2 \mathrm{kHz})$.
In addition to $f$-mode resonances, low-frequency $\mathcal{O}(100 \mathrm{Hz}$--$400 \mathrm{Hz})$ $g$-mode or $i$-mode resonances can also be excited during the inspiral, due to thermal or chemical gradients inside the neutron star~\cite{Lai:1993di,Yu_2016,Gittins:2024oeh,Counsell_2024} or phase transitions~\cite{Counsell:2025hcv}, respectively.
We summarize the different modes that could be excited during a binary neutron star inspiral in Table~\ref{tab:resonances}.

The imaginary part of the tidal response function, the dissipative tidal response, models how dissipative processes within the star introduce a time delay (or ``lag'') in the tidal response. 
In a low-frequency (Taylor) expansion, tidal dissipation is commonly parameterized by a constant, called the dissipative tidal deformability $\Xi$~\cite{Ripley:2023qxo, Poisson:1994yf, Tagoshi:1997jy}.
A first search for tidal dissipation using the GW170817 event has set bounds on the individual $\Xi$ for each star. Follow-up analysis, including next-to-leading-order PN corrections, refined the resulting constraints \cite{Ripley:2023lsq, HegadeKR:2024agt, HegadeKR:2024slr}. 
We present a summary of the different terminologies for the tidal deformabilities used throughout this paper in Table~\ref{tab:tidal-deformabilities}. 

\begin{table}[b]
\begin{center}
\begin{tabular}{| c | c | c | } 
\hline
\T
\textbf{Mode} & $f_{\rm char}$ 
& \textbf{Cause}
\B \\
\hline
\T
$f$-mode & $\mathcal{O}(2 \mathrm{k Hz})$ & Mean density oscillations~\cite{Kokkotas:1999bd}.
\B \\
\hline
\T
$g$-mode & $\mathcal{O}(100 \mathrm{Hz})$ & Chemical composition gradients~\cite{1992ApJ...395..240R}.
\B \\
\hline
\T
$i$-mode & $\mathcal{O}(600 \mathrm{Hz})$ & First-order phase transitions~\cite{Counsell_2024}.
\B \\
\hline
\end{tabular}
\caption{\label{tab:resonances} 
Different modes that could be excited inside a neutron star during the inspiral and the rough characteristic (gravitational-wave) frequency $f_{\rm char}$ at which they are believed to occur.  
}
\end{center}
\end{table}
\begin{table}[t]
\begin{center}
\begin{tabular}{| c | c | } 
\hline
\T
\textbf{Name} & \textbf{Symbol} 
\B
\\
\hline
\T
$f$-dependent tidal response & $\hat{K}_2(f)$ 
\B
\\
\hline
\T
$f$-dependent cons.~tidal deformability & $\Lambda(f) = \frac{2 \mathrm{Re}\left[\hat{K}_2(f)\right]}{3 C^5} $ 
\B
\\
\hline
\T
$f$-dependent diss.~tidal deformability & $\Xi (f) = \frac{2 \mathrm{Im}\left[\hat{K}_2(f)\right]}{6 \pi f C^6 R } $  
\B
\\
\hline
\T
Static love number & $k_2 = \hat{K}_2(f = 0)$ 
\B \\
\hline
\T
Static tidal deformability & $\Lambda = \Lambda(f = 0)$  
\B \\ 
\hline
\end{tabular}
\caption{\label{tab:tidal-deformabilities} 
Definitions of the different static and frequency-dependent tidal deformabilities used throughout the paper. In the table above, $C=M/R$ is the compactness, where $M$ is the mass and $R$ is the radius of the neutron star. 
}
\end{center}
\end{table}
A well-established formalism to model the static tidal deformability $\Lambda$ exists~\cite{Hinderer:2007mb,Binnington:2009bb,Damour:2009vw} in general relativity, allowing one to explore how different nuclear physics properties impact the time-independent tidal deformation of a neutron star.
Extending this approach to frequency-dependent tidal deformations is an active area of research.
Several studies~\cite{Hinderer:2016eia,Steinhoff:2016rfi} have used a \textit{phenomenological} approach to model the frequency-dependent tidal response function by adopting a Newtonian-like mode approximation in effective one-body models, which are calibrated to numerical relativity simulations.
While successfully applied in waveform modeling~\cite{Abac_2024}, the phenomenological approach employed prevents one from understanding the errors associated with the approximation and the calibration.
In addition to this phenomenological approach, several studies use techniques from scattering amplitudes and black hole perturbation theory to model the time-dependent response~\cite{Saketh:2024juq,Miao:2025utd}.

Alternatively, one can use tools from matched asymptotic expansions and PN theory~\cite{Poisson:2020vap} to model the frequency-dependent tidal response function of a neutron star.
In this approach, one can understand how the PN dynamics of the orbit affects the tidal deformations of the binary stars, and systematically improve the approximation by including higher PN order corrections.
This approach has been shown to capture several features in the tidal response function present in Newtonian theory, such as mode resonances~\cite{Pitre:2023xsr,HegadeKR:2024agt}, tidal dissipation~\cite{Ripley:2023lsq,Ripley:2023qxo,HegadeKR:2024agt}, mode expansions~\cite{HegadeKR:2025qwj,Martinez-Rodriguez:2026omk}, and nonlinear tidal interactions~\cite{Pitre:2025qdf}.
Such progress in modeling motivates quantifying, within a single relativistic framework, how microphysical input controls the complex, frequency-dependent tidal response of neutron stars.

Certain works have explored aspects of the interplay between microphysical input and the frequency-dependent tidal response.
\textit{Newtonian models} of tidal interaction have systematically studied how different EoS models can impact the conservative tidal response~\cite{Yu_2016,Passamonti:2022yqp,Andersson:2019dwg} and understand how effective bulk-viscosity approximations can impact the gravitational-wave observables~\cite{Arras:2018fxj,Ghosh:2023vrx,Ghosh:2025wfx}.
Relativistic calculations have also explored the impact of symmetry energy coefficients  (e.g.~$S_{\mathrm{sym}},L_{\mathrm{sym}}$) on neutron-star radii~\cite{Ozel:2016oaf, Lattimer:2000nx, LIGOScientific:2018cki, Most:2018hfd, Capano:2019eae, Miller:2019cac, Dittmann:2024mbo, Raaijmakers:2021uju}, time-independent conservative tidal deformabilities, and $f$-mode frequencies (see~\cite{Lattimer:2000nx, Yagi:2013awa, Yagi:2015pkc, De:2018uhw, Raithel:2018ncd} and references therein).
More recently, binary neutron star simulations~\cite{Most:2021ktk, Chesler:2019osn, Raduta:2022elz} have also explored the impact of $L_{\mathrm{sym}}$ on the post-merger dynamics and found that the amount of dynamical ejecta increases with $L_{\mathrm{sym}}$, while gravitational-wave emission is mostly insensitive to variations of this quantity. To our knowledge, the imprint of the microphysical properties of the nuclear EoS on general-relativistic tidal observables in the inspiral has not previously been investigated self-consistently.  

The key scientific questions we are interested in exploring in this paper are the following:
\begin{itemize}
    \item[i)] How do the nuclear physics parameters impact the conservative tidal response function? And conversely, can a gravitational-wave measurement of the conservative tidal response inform nuclear physics?
    \item[ii)] What parameters control the magnitude of the dissipative tidal response function, and in particular, how does the latter scale with the bulk viscosity computed in linear response calculations?
    \item[iii)] Does the predicted value of the dissipative tidal response function leave a measurable imprint on the gravitational-wave signal?
\end{itemize}
In this paper, we go beyond the previous studies by exploring the frequency-dependent tidal response function in full general relativity, focusing on the quasi-circular inspiral of binary neutron star coalescence.
We therefore compute the full quadrupolar tidal response in general relativity for two representative classes of EoS: a nucleonic (npe) meta-model and a simple three-flavor quark-matter model. The nucleonic EoS assumes matter composed of neutrons, protons, and electrons, and we parameterize it by the temperature $T$ and the symmetry-energy coefficients $S_{\mathrm{sym}}$ (value at saturation), $L_{\mathrm{sym}}$ (slope), and $K_{\mathrm{sym}}$ (curvature) \cite{Li:2019xxz}. For the quark-matter EoS, we adopt an MIT bag model \cite{Johnson:1975zp}. 

In addition to the EoS models, we also include, to our knowledge for the first time, the weak-interaction–driven bulk-viscous effects due to Urca processes and, when in quark matter, strangeness-changing reactions  which can dominate in neutrino-transparent matter during the early inspiral and mergers \cite{Sawyer:1989dp, Haensel:1992zz, Madsen:1992sx, Sad:2007afd, Alford:2017rxf}. Importantly, weak-interaction-driven bulk viscosity does not only enter the dissipative tidal response: in a causal relaxation description, the bulk stress relaxes on a finite time scale, so the induced out-of-equilibrium pressure has both an out-of-phase component (dissipation) and an in-phase component (reactive response). Consequently, the same transport processes can contribute to the conservative dynamical tides in addition to the dissipative part \cite{Yang:2025yoo}. Therefore, we include the reaction rates directly in our model through an evolution equation for the species fraction~\cite{Most:2022yhe} and do not adopt effective bulk-viscosity approximations, which were commonly used in previous studies~\cite{Most:2021zvc,HegadeKR:2024agt,Saketh:2024juq, Chabanov:2023blf, Chabanov:2023abq}.

For question (i), our results indicate that, for $npe$ matter, the conservative tidal response is affected by an effective combination of $(L_{\mathrm{sym}}, K_{\mathrm{sym}})$, with the largest variation along the $K_{\mathrm{sym}}$ direction. This conclusion is, in part, due to the lack of stringent constraints on $K_{\mathrm{sym}}$ from nuclear physics experiments~\cite{Li:2019xxz, Xie:2020tdo}.
However, the conservative tidal response is more sensitive to variations in $L_{\mathrm{sym}}$ in the low mass regime where we expect to measure tidal signatures.
This sensitivity indicates that the dynamical tidal response carries information on both $K_{\mathrm{sym}}$ and $L_{\mathrm{sym}}$. Exploiting it as a quantitative constraint requires a dedicated parameter-estimation study, which we leave to future work.
We also find that the $g$-mode frequencies are sensitive to $L_{\mathrm{sym}}$, increasing systematically toward the stiff, large-$L_{\mathrm{sym}}$ end of the currently allowed range (illustrated here using PREX-II-like values).
For the quark model, the only parameter we vary is the bag constant and we find that conservative tidal response shows a strong variation with bag constant.

For questions (ii) and (iii), we find that the dissipative tidal deformability is highly sensitive to the temperature of the star and it scales almost linearly with the bulk viscosity until a relaxation peak is approached. However, we find that the predicted value of the dissipative tidal deformability is always a few orders of magnitude smaller than what is potentially observable by third-generation gravitational-wave detectors~\cite{Ripley:2023lsq,HegadeKR:2024slr}.
This suggests that bulk viscosity alone will not lead to a sufficiently large dissipative tidal deformability to be observable.
Other micro-physical channels of bulk-viscous-like dissipation (such as hyperonic degrees of freedom~\cite{Ghosh:2023vrx}, shear viscosity~\cite{Saketh:2024juq,Shterenberg:2024tmo}, or effective sources, such as turbulence, or critical slowing down near a possible QCD critical point~\cite{Karthein:2026rok}), however, may lead to larger dissipative tidal deformabilities, and therefore, these need to be studied further. 
We emphasize that the purpose of the dissipative-tide analysis is twofold: to present the full complex tidal response within a single relativistic framework, and to establish quantitatively whether weak-interaction-driven bulk viscosity---the minimal dissipative channel considered here---leaves an observable imprint. Our finding that it does not is an informative null result, and is distinct from our use of the conservative dynamical tide as a probe of dense matter.
We also note that previous claims of measurability have relied on results from Newtonian physics~\cite{Ghosh:2025wfx} and those results produce systematically larger values for the dissipative tidal deformability than what is seen in our fully general-relativistic work.

The rest of the paper explains these results in detail and is organized as follows.
In Sec.~\ref{sec:EOS-models}, we introduce analytical EoS models and describe our parameterization in terms of nuclear-symmetry-energy observables.
Section~\ref{sec:tidal-response-review} reviews the formalism to calculate the tidal response function in general relativity.
In Secs.~\ref{sec:results} and~\ref{sec:conclusions}, we present our results and conclusions respectively. 
Finally, we present the calculations of thermodynamic derivatives and weak interaction rates in the Appendices.

\noindent
\emph{Notation}: We use geometric units in this paper and set $c=1=G=k_B$.
The signature of the metric is  $(-,+,+,+)$. 
\section{Microphysical modeling}\label{sec:EOS-models}
Our system (regardless of the choice made about the EoS) can be described by a quasi-equilibrium thermodynamic state with an entropy density, baryon density, and charge fraction $\{s, n_B, Y_X\}$, respectively, where $Y_X=n_X/n_B$ is not necessarily in electroweak equilibrium, and $X$ denotes a given conserved charge, i.e., $X=Q,S$ for electric charge and strangeness. 
The system can then be characterized by the equations describing entropy production, baryon number conservation, and the dynamics of the charge current \cite{Gavassino:2020kwo, Celora:2022nbp,Gavassino:2023xkt},
\begin{subequations}\label{conservation}
\begin{align}
& \nabla_\mu s^\mu = \frac{Q_{\nu_e}}{T} + \frac{\delta \mu_X \,\Gamma_X}{T} \,, \\
&\nabla_\mu (n_B u^\mu) = 0 \,, \\ 
&u^\mu \nabla_\mu Y_X = \frac{\Gamma_X}{n_B} \,,
\end{align}
\end{subequations}
where $Q_{\nu_e}$ stems from the (assumed isotropic) radiative energy loss due to neutrinos, and $\Ga_X$ is the flavor equilibration rate for the corresponding charge fraction $Y_X$.

We then consider two different classes of EoS: nucleonic ($npe$ matter) and 3-flavor quark matter. 
\begin{itemize}
    \item In the case of $npe$ matter, we have no net strangeness, so we only have to take the electric charge into account. 
    The star must still be electrically neutral, such that $n_p=n_e$, and then our electric charge fraction is $Y_Q=n_e/n_B=n_p/n_B$. 
    However, a chemical imbalance can occur out of \emph{electroweak} equilibrium such that $\delta\mu_Q = \mu_n - \mu_p - \mu_e\neq 0$, which is dictated by the time delay to approach equilibrium. 
\item In the case of quark matter, we may have both a net strangeness and a net electric charge such that $X=\left\{Q,S\right\}$. The primary electroweak equilibrium reaction for electric charge is dictated by $d\rightarrow u+e^-+\bar{\nu}_e$ since $Q_d=-1/3$ and $Q_u=+2/3$. 
However, strange quark decays contribute to electric charge as well as strangeness flavor-changing processes through $s\rightarrow u+e^-+\bar{\nu}_e$ since $Q_s=-1/3$, $S_s=-1$, and $S_u=S_d=0$.  Then, one would have two different chemical potential imbalances to consider: $\delta\mu_Q = \mu_d - \mu_u - \mu_e$ and $\delta\mu_S = \mu_s - \mu_d$. 
In principle, one should track both $\delta\mu_Q$ and $\delta\mu_S$, but we have checked that the influence of $\delta\mu_Q$ is negligible in the quark matter case and, thus, we only track $\delta\mu_S$.
\end{itemize}
For both systems, electroweak equilibrium is achieved exactly at $\de\mu_S = \de\mu_Q =0$.

The microscopic degrees of freedom are coupled to gravity through the Einstein equations and the stress-energy conservation equations\footnote{Strictly, the Einstein equations $G_{\mu\nu}=8\pi T_{\mu\nu}^{\rm total}$ and the Bianchi identity imply conservation of the \emph{total} stress-energy tensor $T_{\mu\nu}^{\rm total}=T_{\mu\nu}+T_{\mu\nu}^{(\nu)}$, which includes the neutrino contribution $T_{\mu\nu}^{(\nu)}$. Eq.~(3) then follows from $\nabla_\mu T_{\rm total}^{\mu\alpha}=0$ under the assumption of isotropic neutrino free-streaming. In the inspiral regime considered here, $Q_{\nu_e}\approx 0$, so this distinction does not affect our results.},
\begin{align}
    &G_{\mu \nu} = 8 \pi T_{\mu \nu} \,,\\
    &\nabla_{\mu} T^{\mu \alpha} = - Q_{\nu_e} u^\alpha\,,
\end{align}
At the macroscopic level, we model both $npe$ matter and quark matter as an effective fluid whose stress-energy tensor has the algebraic form of an ideal fluid,
\begin{align}
    T_{\mu \nu} = (\varepsilon + P) u_\mu u_\nu + P g_{\mu \nu}\,,
\end{align}
where $\varepsilon(s, n_B, Y_X)$ is the energy density, and $P(s, n_B, Y_X)$ is the pressure. However, the fluid is not strictly perfect, and the bulk-viscous effects are incorporated through a viscous correction to the pressure~\cite{Gavassino:2020kwo, Celora:2022nbp, Gavassino:2023xkt}. Concretely, we write, 
\begin{equation}
    P = P(s, n_B, Y_X=Y_X^{\mathrm{eq}}) + \Pi(s, n_B, Y_X)\,,
\end{equation} where $Y_X^{\mathrm{eq}}$ is the $\beta$-equilibrium charge fraction, and $\Pi$ is the bulk scalar induced by chemical imbalance, which characterizes the deviation from the equilibrium pressure and vanishes when $Y_X=Y_X^{\mathrm{eq}}$\footnote{We note that this parameterization coincides with the Landau frame at linear order in $\delta\mu_X$. Since the tidal perturbation equations in Sec.~\ref{sec:tidal_perturb} are also linearized in the perturbation amplitude, and $\delta\mu_X$ is itself first order in this amplitude, the distinction between frames does not affect our results.}.

During the inspiral phase of binary neutron stars, when the tidal deformation is small, the system may be appropriately described by the leading-order, frequency-dependent pressure response to a small volume change. Therefore, we truncate the expressions at linear order and obtain the following approximations: $Q_{\nu_e} \sim Q_{\nu_e}^0 + \mathcal{O}(\delta \mu^2)$ 
and $\Ga_X \sim \kappa\,\de\mu_X$, 
where $Q_{\nu_e}^0 \approx 0$ for typical temperatures associated with the inspiral phase \cite{Iwamoto:1982zz, Adhya:2016wql, Camelio:2022ljs} and the form of $\kappa$ is detailed in Appendices~\ref{appendix:Urca} and \ref{appendix:quark_rates}. \footnote{We note that the linearization $\Gamma_X \sim \kappa\,\delta\mu_X$ is tied to the linearization in the tidal perturbation amplitude: since $\delta\mu_X$ is itself first order in the tidal amplitude $h$ (it is sourced by the tidally induced oscillations of the composition), the two linearizations are mutually consistent within linear perturbation theory. As the inspiral progresses and the tidal amplitude grows, however, $\delta\mu_X$ may become comparable to, or exceed, $T$, and the system can enter the suprathermal regime where $\Gamma_X$ is no longer linear in $\delta\mu_X$; our treatment therefore becomes less accurate in this late-inspiral, low-temperature limit, and a proper treatment would require retaining the full nonlinear structure of $\Gamma_X(\delta\mu_X)$.}
With these assumptions, Eq.~\eqref{conservation} simplifies to
\begin{subequations}\label{eq:conservation-sim}
\begin{align}
& \nabla_\mu s^\mu = 0 \,, \\
&\nabla_\mu (n_B u^\mu) = 0 \,, \\ 
&u^\mu \nabla_\mu Y_X = \frac{\kappa\,\delta\mu_X}{n_B} \,.
\end{align}
\end{subequations}

The rest of this section explains our EoS models in detail. Section~\ref{sec:NPE-EOS} describes the $npe$ model. The EoS model for the quark EoS is presented in Sec .~\ref {sec:quark-EOS}.
Section~\ref{sec:bulk-viscosity} describes the calculation of the bulk viscosity for the different EoS models studied in this work. 
\subsection{$npe$ Equation of State}\label{sec:NPE-EOS}
In order to compute the bulk viscous effects from the inspiral, we consider a simple analytical model (often called a meta-model) that is written in terms of parameters
such as symmetry-energy coefficients and nuclear saturation properties. In this model, neutron stars are composed of $npe$ matter: neutrons $n$, protons $p$, and electrons $e$.\footnote{We neglect muons throughout this analysis. At densities relevant for the neutron-star core, the electron chemical potential generically exceeds the muon rest mass, so muons are present and contribute an additional equilibration channel that modifies the bulk viscosity both qualitatively and quantitatively, including the possibility of a two-peak structure in $\zeta(T)$~\cite{Haensel:2000vz,Alford:2023gxq,Hernandez:2026tak}. Incorporating muons into our meta-model and studying how their contribution propagates into the conservative and dissipative tidal response is a natural extension of this work.} Specifically, we adopt the parabolic approximation for the nuclear symmetry energy to describe nuclear matter \cite{PhysRevC.44.1892}, where the energy per baryon $E$ is expanded in terms of the isospin asymmetry term\footnote{Note this is only valid in the absence of strangeness, otherwise isospin asymmetry must be defined as $\delta_I=1+Y_S-2Y_Q$, see \cite{Yang:2025wop, Danhoni:2025qpn}.} $\delta_Q\equiv 1-2Y_Q$
 up to second order,
\begin{equation} \label{eq:PA}
    E(n_B, \delta_Q) \approx E(n_B, 0) + E_{\mathrm{sym}}(n_B)\, \delta_Q^2 \, , 
\end{equation}
where the nuclear symmetry energy $E_{\mathrm{sym}}(n_B)\approx E(n_B,1) -E(n_B,0) $ is approximately the energy difference between pure neutron matter and symmetric nuclear matter (where these limits are only well defined for $npe$). 
Within the parabolic approximation, it is also customary to perform a second expansion where $E_{\mathrm{sym}}(n_B)$ is expanded around the saturation density $n_{\mathrm{sat}}$. Here, we truncate the expansion at second order to isolate the impact of the better-constrained symmetry-energy coefficients while keeping the parameter scan minimal; this truncation has been shown to work quite well for chiral effective field theory \cite{Wellenhofer:2015qba, Wellenhofer:2016lnl, Drischler:2020hwi, Drischler:2021kxf} and a chiral mean field model \cite{Yang:2025wop}:
\begin{align}\label{eq:Esym_quad}
    E_{\mathrm{sym}}(n_B) \approx&\, S_{\mathrm{sym}} + L_{\mathrm{sym}}\left( \frac{n_B - n_{\mathrm{sat}}}{3 n_{\mathrm{sat}}}\right) \nonumber \\
    &+ \frac{K_{\mathrm{sym}}}{2} \left( \frac{n_B - n_{\mathrm{sat}}}{3 n_{\mathrm{sat}}}\right)^2 \, ,
\end{align}
where $S_{\mathrm{sym}}$ is the nuclear symmetry energy at $n_{\mathrm{sat}}$, $L_{\mathrm{sym}}$ is the first-order coefficient (or slope), and $K_{\mathrm{sym}}$ is the second-order coefficient (or curvature). 
The expansion in Eq.~(8) is the standard meta-model parametrization of nucleonic matter~\cite{Margueron:2017eqc, Margueron:2017lup}, which reproduces realistic Skyrme and relativistic-mean-field equations of state to good accuracy and, consistent with our findings, identifies $L_{\rm sym}$ and $K_{\rm sym}$ as the most influential isovector parameters~\cite{Margueron:2017eqc, Dutra:2014qga}. We also note that the second-order density truncation is anchored near saturation, so at the few-times-saturation densities reached in the core the onset of new degrees of freedom (e.g., hyperons or a quark-matter phase) would require extending it; the corresponding generalization to strangeness-bearing matter, validated against microscopic models and perturbative QCD, is developed in Refs.~\cite{Yang:2025wop, Danhoni:2025qpn}.

The ranges for the symmetry-energy parameters in our model are guided by nuclear experimental and theoretical considerations, including nuclear mass fitting \cite{Dutra:2012mb, Dutra:2014qga, Tagami:2022xvs, Kortelainen:2010hv}, neutron skin thickness measurements \cite{Steiner:2004fi, Reed:2021nqk, Brown:2000pd, Typel:2001lcw, Xu:2020fdc, Zhang:2013wna}, chiral effective field theory, and ab-initio calculations \cite{Drischler:2020hwi, Tews:2024owl}. 
Although the symmetry energy at saturation density, $S_{\mathrm{sym}}$, is relatively well constrained experimentally, the slope $L_{\mathrm{sym}}$ is less constrained from experiment, and higher-order coefficients have only theoretical constraints. In particular, the parity-violating electron-scattering measurements of the neutron skins of ${}^{208}$Pb (PREX and PREX-II) \cite{PREX:2021umo} and ${}^{48}$Ca (CREX) \cite{CREX:2022kgg} have been interpreted as favoring different values of $L_{\mathrm{sym}}$---a comparatively large value from the ${}^{208}$Pb data and a comparatively small one from the ${}^{48}$Ca data---and the two have not yet been reconciled within a single model \cite{Lattimer:2023rpe}. Considerable theoretical effort has been devoted to understanding this tension \cite{Furnstahl:2001un, Danielewicz:2013upa, Zhang:2022bni, Reinhard:2022inh, Lattimer:2023rpe}. Guided by the compilation in \cite{MUSES:2023hyz}, we scan $S_{\mathrm{sym}} \in [29, 34]~\mathrm{MeV}$ and $L_{\mathrm{sym}}\in[40,120]~\mathrm{MeV}$ in this work; this range covers the values most commonly adopted in the literature, including the stiff isovector sector favored by PREX-II, while the lowest $L_{\mathrm{sym}}$ values inferred from CREX lie below it.

As in the case of symmetric nuclear matter ($\delta_Q=0$), we model the energy per baryon with a Pad\'e-like ansatz. This provides a flexible global parametrization over a broad density range with relatively few coefficients, while also making it straightforward to enforce sensible asymptotic behavior—e.g., a speed of sound that saturates toward a chosen value at high density:
\begin{align}
    E(n_B,0) &= m_B \frac{\sum_{i=1}^3 a_i\,(n_B/n_{\mathrm{sat}})^i}{1+\sum_{i=1}^3 b_i\,(n_B/n_{\mathrm{sat}})^i} + E_b \, ,
\end{align}
where $a_i$ and $b_i$ are dimensionless fit parameters, $m_B=938.9~\mathrm{MeV}$ is the averaged vacuum nucleon mass, and $n_{\mathrm{sat}}=0.17~\mathrm{fm}^{-3}$ is the saturation density. The constant $E_b$ is the binding energy per baryon at saturation, defined by
\begin{align}\label{eq:Eb_def}
    E_b \equiv E(n_{\mathrm{sat}},0)-m_B \, .
\end{align}

To connect the ansatz to empirical saturation properties, it is convenient to expand $E(n_B,0)$ about $n_{\mathrm{sat}}$,
\begin{align}\label{eq:Esym_Taylor}
E(n_B,0)=&E(n_{\mathrm{sat}},0)+\left.\frac{dE(n_B,0)}{dn_B}\right|_{n_{\mathrm{sat}}}(n_B-n_{\mathrm{sat}}) \nonumber \\
&+\frac12\left.\frac{d^2E(n_B,0)}{dn_B^2}\right|_{n_{\mathrm{sat}}}(n_B-n_{\mathrm{sat}})^2 \nonumber \\
&+ \mathcal{O}\!\left[(n_B-n_{\mathrm{sat}})^3\right].
\end{align}
Saturation means that symmetric matter has vanishing pressure at $n_{\mathrm{sat}}$, which is equivalent to the stationarity condition
\begin{align}\label{eq:sat_stationary}
    \left.\frac{d}{dn_B}E(n_B,0)\right|_{n_{\mathrm{sat}}}=0 \, ,
\end{align}
so the linear term in Eq.~\eqref{eq:Esym_Taylor} vanishes. The curvature at saturation is encoded in the incompressibility (compression modulus)
\begin{align}\label{eq:K0_def}
    K_0 \equiv 9 n_{\mathrm{sat}}^2 \left.\frac{d^2E(n_B,0)}{dn_B^2}\right|_{n_{\mathrm{sat}}} \, ,
\end{align}
which measures the quadratic stiffness of symmetric matter against uniform compression/rarefaction around $n_{\mathrm{sat}}$. Using Eqs.~\eqref{eq:Eb_def}--\eqref{eq:K0_def}, the expansion becomes
\begin{align}\label{eq:sym_matter}
E(n_B,0)=&m_B + E_b +\frac{K_0}{18}\left(\frac{n_B-n_{\mathrm{sat}}}{n_{\mathrm{sat}}}\right)^2 \nonumber \\
& + \mathcal{O}\!\left[(n_B-n_{\mathrm{sat}})^3\right].
\end{align}

To fix the parameters in the Padé-like ansatz, we select representative values for parameters $(b_1,b_2,b_3)$ and enforce the three saturation constraints \eqref{eq:Eb_def}, \eqref{eq:sat_stationary}, and \eqref{eq:K0_def} to determine  parameters $(a_1,a_2,a_3)$,   while ensuring that the resulting EoS is thermodynamically stable, exhibits a causal speed of sound at all densities, and  is qualitatively similar to EoSs obtained from standard relativistic mean field models.

We fix the binding energy at saturation to $E_b = -16.3~\mathrm{MeV}$\cite{Myers:1966zz} and adopt $K_0 = 250~\mathrm{MeV}$ \cite{Colo:2013yta, Todd-Rutel:2005yzo, Colo:2004mj, Agrawal:2003xb, Khan:2012ps, Stone:2014wza} as a reference value, both consistent with current constraints from nuclear structure and heavy-ion data. Although $K_0$ enters the fitting procedure that determines $(a_1,a_2,a_3)$, it is not held fixed throughout the analysis: in later sections we vary $K_0$ within its empirical range and re-solve for the Padé coefficients at each value. Since our goal is to isolate the qualitative impact of varying the symmetry-energy sector on the tidal response, we do not attempt a full statistical propagation of the uncertainties in $E_b$, $K_0$, and related saturation properties in this work. For the reference value $K_0 = 250~\mathrm{MeV}$, the resulting coefficients are $(a_1,a_2,a_3,b_1,b_2,b_3) = (0.73,-1.46,0.73,42,6,0.25)$.

In addition, we include temperature and lepton contributions to the total energy density to account for weak interactions and tidal heating. We emphasize that the temperature is the local temperature measured in the fluid rest frame; we neglect thermal transport and treat the temperature profile as prescribed:
\begin{align}\label{eq:model}
    \varepsilon(n_B, \delta_Q, T) 
    =& n_B\;E(n_B, 0) + n_B\;E_{\mathrm{sym}}(n_B)\;  \delta_Q^2\nonumber \\ 
    &+ \frac{n_B}{4m_B} T^2 + \varepsilon_e(n_B,  Y_Q, T) \, ,
\end{align}
where $\varepsilon_e$ and $n_e = n_B\,Y_Q$ are computed as a free Fermi gas,
\begin{subequations}
    \begin{align}
        \varepsilon_e &= \frac{1}{\pi^2\hbar^3} \int^\infty_0 dk\; \frac{k^2 (k^2 + m_e^2)^{1/2}}{e^{(\sqrt{k^2 + m_e^2}-\mu_e)/T} + 1} \, , \\
        n_e &= \frac{1}{\pi^2\hbar^3} \int^\infty_0 dk\; \frac{k^2}{e^{(\sqrt{k^2 + m_e^2}-\mu_e)/T} + 1} \, .
    \end{align}
\end{subequations}
To keep the model simple and analytic, we assume the electron mass $m_e = 0$, and adopt the Sommerfeld expansion
\cite{chandrasekhar1957introduction},
\begin{widetext}
\begin{subequations}
    \begin{align}
        \varepsilon_e &= \frac{1}{\pi^2\hbar^3}\left[ \frac{\mu_e^4}{4} + \frac{\pi^2}{2} (\mu_e T)^2 \right] \, , \\
        \mu_e &= 
        \left( \frac{3}{2}\pi^2\hbar^3 n_B\, Y_Q + \sqrt{\frac{1}{4}\left( 3\pi^2\hbar^3 n_B\,Y_Q \right)^2 + \frac{1}{27} (\pi T)^6} \right)^{1/3}  - \frac{(\pi T)^2}{3 \left( \frac{3}{2}\pi^2\hbar^3 n_B\,Y_Q + \sqrt{\frac{1}{4}\left( 3\pi^2\hbar^3 n_B\, Y_Q \right)^2 + \frac{1}{27} (\pi T)^6} \right)^{1/3}}.
    \end{align}
\end{subequations}
\end{widetext}
Thermodynamic stability of Eq.~\eqref{eq:model} can be established by verifying that the Hessian of the energy density is positive definite \cite{landau2013statistical},
\begin{equation}
    H_{AB}=
    \begin{pmatrix}
        \left.\dfrac{\partial T}{\partial s}\right|_{n_B,n_X} &
        \left.\dfrac{\partial T}{\partial n_B}\right|_{s,n_X} &
        \left.\dfrac{\partial T}{\partial n_X}\right|_{s,n_B} \\
        \left.\dfrac{\partial \mu_B}{\partial s}\right|_{n_B,n_X} &
        \left.\dfrac{\partial \mu_B}{\partial n_B}\right|_{s,n_X} &
        \left.\dfrac{\partial \mu_B}{\partial n_X}\right|_{s,n_B} \\
        \left.\dfrac{\partial \mu_X}{\partial s}\right|_{n_B,n_X} &
        \left.\dfrac{\partial \mu_X}{\partial n_B}\right|_{s,n_X} &
        \left.\dfrac{\partial \mu_X}{\partial n_X}\right|_{s,n_B}
    \end{pmatrix}.
\end{equation}
For all parameter sets used in this study, we verified that the thermodynamic consistency and stability relations are satisfied. For a given state $(n_B,T,\delta_Q)$, we build the pressure as a function of these variables through the usual Legendre relation 
\begin{equation}
  P(n_B,T,\delta_Q)
  = n_B^2 \left.\frac{\partial}{\partial n_B}
    \left(\frac{\varepsilon(n_B,T,\delta_Q)}{n_B}\right)\right|_{s/n_B,Y_Q}.
\end{equation}
where $\varepsilon$ is given in Eq.~\eqref{eq:model}. This then defines our nuclear-matter EoS.

In this study, we construct a hybrid EoS as follows. For baryon densities $n_B<0.85\,n_{\mathrm{sat}}$ we adopt the Togashi EoS \cite{Togashi:2017mjp}, which we use in its cold ($T=0$) form and therefore keep fixed (we do not include temperature dependence in this low-density segment). For $n_B>n_{\mathrm{sat}}$, we use our analytic finite-temperature $npe$ EoS. In the transition region $0.85\,n_{\mathrm{sat}}\le n_B\le n_{\mathrm{sat}}$ we construct a smooth matching by interpolating the thermodynamic quantities as functions of $n_B$, with the interpolation chosen so that $P(n_B)$ is continuous and differentiable at $n_B=0.85\,n_{\mathrm{sat}}$ and $n_B=n_{\mathrm{sat}}$. Since the bulk-viscous dissipation is negligible in the low-density region \cite{Yang:2025yoo}, our qualitative results are insensitive to the detailed choice of matching scheme near the crust--core interface and to neglecting finite-$T$ effects in the Togashi segment.

\subsection{Quark Equation of State}\label{sec:quark-EOS}
As an alternative to a typical neutron star with a crust and $npe$ at its core, we also explore a hybrid configuration with a three-flavor quark-matter core and a hadronic outer layer.
For the quark EoS, we use a toy model for quarks, the MIT bag model \cite{Johnson:1975zp}. The model assumes that the baryon is composed of three non-interacting quarks inside a bag, and that all information about the strong force is encoded in the bag constant, $B$. For each flavor of quarks, the number density, pressure, and energy density take the form of free fermions. Taking the zero-temperature approximation,
\begin{subequations}
    \begin{align}
        n_f &= \frac{k_f^3}{\pi^2}\,, \\
        P_f &= \frac{1}{4\pi^2} \left[ \mu_f k_f \left( \mu_f^2 - \frac{5}{2} m_f^2 \right) + \frac{3}{2} m_f^4 \ln\left( \frac{\mu_f + k_f}{m_f} \right) \right]\,, \\
        \varepsilon_f &= \frac{3}{4\pi^2}\left[\mu_f k_f\big(\mu_f^2-\tfrac12 m_f^2\big)-\tfrac12 m_f^4\ln\frac{\mu_f+k_f}{m_f}\right]
    \end{align}
\end{subequations}
where the subscript $f$ denotes the flavor, $k_f$ is the Fermi momentum, and $\mu_f = \sqrt{m_f^2 + k_f^2}$ is the Fermi chemical potential. Here, we only consider the three light quarks: up, down, and strange quarks. The total pressure and energy density are then,
\begin{subequations}\label{eq:quark-energy-and-number-density}
    \begin{align}
        \varepsilon &= \sum_{f = u, d, s} \varepsilon_f + \varepsilon_e + B, \\
        P &= \sum_{f = u, d, s} P_f + P_e - B \label{eq:quark-eos}
    \end{align}
\end{subequations}
For simplicity, we set the mass of the up  and down quarks to zero, and only take the zero temperature part of $P_e$ and $\varepsilon_e$. In this work, we treat this bag-model EoS as a controlled toy model that provides a qualitatively different microphysical timescale compared to nucleonic matter. In contrast to the $npe$ meta-model—where we vary isovector (symmetry-energy) parameters while anchoring the isoscalar sector at saturation—here the stiffness and self-binding of quark matter are primarily controlled by the bag constant $B$ (and by the strange-quark mass $m_s$). 

Similar to the $npe$ EoS, we connect the quark-matter core to a hadronic outer layer. We emphasize that the resulting object is a hybrid configuration rather than a self-bound bare strange star: the wide, smooth matching over $0.85\,n_{\mathrm{sat}}\le n_B\le 2\,n_{\mathrm{sat}}$ is a modeling choice that allows a uniform numerical treatment of the stellar surface across both EoS classes, avoiding the dedicated treatment that a bare quark surface, with its finite surface energy density, would otherwise require in the tidal matching.
For baryon densities $n_B<0.85\,n_{\mathrm{sat}}$ we adopt the Togashi EoS \cite{Togashi:2017mjp}, while for $n_B>2n_{\mathrm{sat}}$, we use the MIT bag model. In the transition region $0.85\,n_{\mathrm{sat}}\le n_B\le 2 n_{\mathrm{sat}}$, we construct the same smooth matching as the procedure for the analytical $npe$ model.

We assume electrically neutral matter with electrons and further adopt the standard separation of weak timescales in degenerate quark matter: reactions that change electric charge equilibrate rapidly compared to the non-leptonic strangeness-changing channel that dominates the bulk viscosity. Concretely, we enforce charge neutrality and take the electric-charge chemical imbalance to vanish, $\delta\mu_Q=\mu_d-\mu_u-\mu_e=0$, while allowing departures from strangeness equilibrium through $\delta\mu_S=\mu_s-\mu_d\neq 0$. In this setup, it is convenient to parametrize the non-equilibrium composition by $(n_B,Y_S)$, where $n_B=(n_u+n_d+n_s)/3$ and $Y_S\equiv n_S/n_B$ with $n_S=-n_s$ the strangeness charge density. At fixed $(n_B,Y_S)$ we determine the Fermi momenta by solving the algebraic constraints from $n_B$, $Y_S$, charge neutrality $n_Q=(2/3)n_u-(1/3)(n_d+n_s)-n_e=0$, and $\delta\mu_Q=0$, and then evaluate $P$ and $\varepsilon$ from Eq.~\eqref{eq:quark-energy-and-number-density}.

Finally, we keep the background EoS at $T=0$ (and include temperature only through the weak rates entering $\Gamma_S$) because the inspiral regime of interest is highly degenerate, $T\ll \mu_f$, so thermal corrections to the bulk thermodynamics are parametrically suppressed compared to the dominant temperature dependence of the reaction rates. Note that the assumptions made here would not hold for color superconducting phases, which exhibit nontrivial temperature effects (see, e.g., \cite{Gholami:2025yqf}). These phases can also affect neutrino absorption \cite{Alford:2025jtm}. 

Similar to the case of $npe$ matter, we use the pressure in Eq.~\eqref{eq:quark-eos} to prescribe the quark EoS, where both $\mu_f$ and $k_f$ can be expressed in terms of $n_B$ and $Y_S$, such that $P = P(n_B, Y_S)$. 
While we use the MIT bag model as an initial attempt, the formalism can be combined with modern perturbative quantum chromodynamics constrained high-density EoSs \cite{Kurkela:2009gj, Gorda:2018gpy, Komoltsev:2023zor}.
\subsection{Bulk Viscosity}\label{sec:bulk-viscosity}
For both cases, weak-interaction-induced flavor exchange drives the system out of $\beta$ equilibrium and leads to dissipation. If we assume a small linear perturbation in terms of $\de\mu_X$, we can expand the pressure around $\beta$ equilibrium \cite{Gavassino:2020kwo},
\begin{align}
P(s, n_B, Y_X) = P|_{\delta \mu = 0} + \Pi.
\end{align}
where $\Pi = P_1 \delta \mu_X$, and $P_1 = \left. {\partial P}/{\partial \delta \mu_X}\right|_{s, n_B, \delta \mu_X = 0}$. 

The change in $\de\mu_X$ can be written in terms of the state variables $\{s, n_B, Y_X\}$,
\begin{align}
\label{beta chain}
    u^\mu \nabla_\mu \delta \mu_X =& \frac{\partial \delta \mu_X}{\partial s}\bigg|_{n_B, Y_X} u^\mu \nabla_\mu s \nonumber + \frac{\partial \delta \mu_X}{\partial n_B}\bigg|_{s, Y_X} u^\mu \nabla_\mu n_B \\
    & + \frac{\partial \delta \mu_X}{\partial Y_X}\bigg|_{s, n_B} u^\mu \nabla_\mu Y_X.
\end{align}
Substituting Eq.~\eqref{eq:conservation-sim}
into Eq.~\eqref{beta chain}, we obtain
\begin{align}
u^\mu \nabla_\mu \delta \mu_X =& \frac{\partial \delta \mu_X}{\partial Y_X}\bigg|_{s, n_B} \frac{\kappa}{n_B} \de\mu_X \nonumber \\
& - \nabla_\mu u^\mu\, \left( n_B \left.\frac{\partial \delta \mu_X}{\partial n_B}\right|_{s, Y_X} + s \frac{\partial \delta \mu_X}{\partial s}\bigg|_{n_B, Y_X} \right),
\end{align}
where we only keep terms up to linear order in  $\de\mu_X$. 
Multiplying this equation by $P_1$ and adding $\de\mu_X\, u^\mu \nabla_\mu P_1$, we find
\begin{align}
\label{delta mu eom}
u^\mu \nabla_\mu \Pi =& \frac{\partial \delta \mu_X}{\partial Y_X}\bigg|_{s, n_B} \frac{\kappa}{n_B} \Pi + \de\mu_X\, u^\mu \nabla_\mu P_1 \nonumber \\
&- \nabla_\mu u^\mu \, P_1 \left( n_B \left.\frac{\partial \delta \mu_X}{\partial n_B}\right|_{s, Y_X} + s \frac{\partial \delta \mu_X}{\partial s}\bigg|_{n_B, Y_X} \right)\,. 
\end{align}
Because $P_1$ is determined at $\delta \mu_X = 0$, it can be written as a function of two state variables,
\begin{equation}
    P_1 = P_1(s, n_B).
\end{equation}
Its derivative along the fluid 4--velocity is 
\begin{equation}
    u^\mu \nabla_\mu P_1 = \frac{\partial P_1}{\partial s} \bigg|_{n_B} u^\mu \nabla_\mu s + \frac{\partial P_1}{\partial n_B}\bigg|_s u^\mu \nabla_\mu n_B.
\end{equation}
If we substitute Eq.~\eqref{eq:conservation-sim}
into the above equation, while keeping only leading-order terms, we have
\begin{equation}
\label{P1 expansion}
    u^\mu \nabla_\mu P_1 =-\frac{\partial P_1}{\partial s} \bigg|_{n_B} s \nabla_\mu u^\mu - \frac{\partial P_1}{\partial n_B}\bigg|_s n_B \nabla_\mu u^\mu.
\end{equation}
\begin{figure}[thp]
    \centering
    \includegraphics[width=0.95\linewidth]{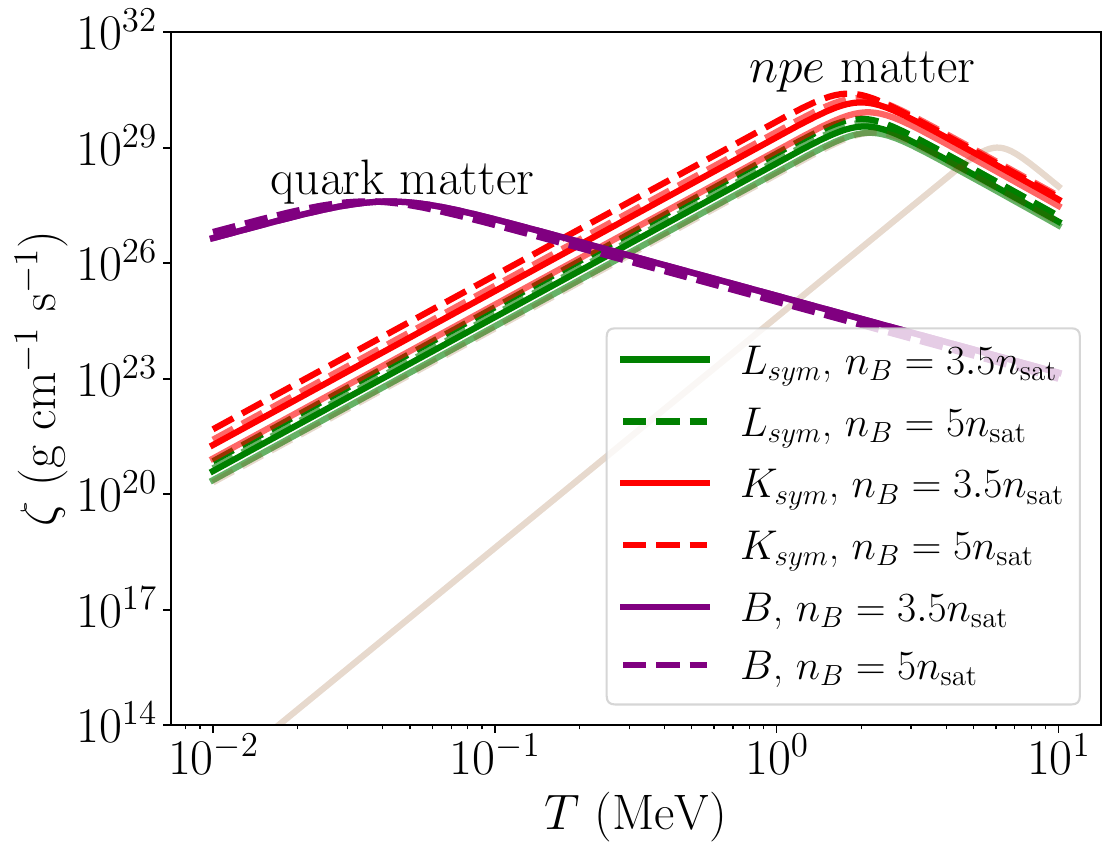}
    \caption{Frequency-dependent bulk viscosity versus temperature, at $\om = 2\pi\times1$ kHz, for various $L_{\mathrm{sym}}$ (green),  $K_{\mathrm{sym}}$ (red), and $B$ (purple), at $n_B=3.5n_{\mathrm{sat}}$ (solid) and $n_B=5n_{\mathrm{sat}}$ (dashed).
    Lighter colored curves indicate lower values of the parameter of interest. 
    In some temperature ranges, two curves (the green and red traces) are indistinguishable and lie on top of each other; the resulting overplotting appears as a light brown trace and does not correspond to an additional parameter choice. For $npe$ matter, increasing either $L_{\mathrm{sym}}$ or $K_{\mathrm{sym}}$---and density---systematically increases the bulk viscosity across most temperatures. In contrast, for quark matter, the bulk viscosity is mostly insensitive to the changes in the bag constant and densities, aside from a modest change for the largest value of $B$ and at $5n_{\mathrm{sat}}$, but its peak occurs at a markedly different temperature than the peak for $npe$ matter.
    } 
    \label{fig:bulk-viscosity-example-plots}
\end{figure}

Finally, we can substitute Eq.~\eqref{P1 expansion} into Eq.~\eqref{delta mu eom} to find
\begin{align}
\label{eq:ISfinalform}
    &u^\mu \nabla_\mu \Pi = - \frac{\Pi}{\tau_\Pi} - \frac{\zeta_0}{\tau_\Pi} \nabla_{\mu} u^\mu - \frac{\de_{\Pi\Pi} \Pi}{\tau_\Pi}\nabla_\mu u^\mu, 
\end{align}
which is the Israel-Stewart equation \cite{Israel:1979wp} with coefficients computed in $\beta$ equilibrium \cite{Gavassino:2020kwo,Gavassino:2023xkt}. We have here defined
\begin{subequations}\label{eq:transport}
    \begin{align}
        &\tau_{\Pi} = - \frac{n_B}{\kappa} \left( \frac{\partial \delta \mu_X}{\partial Y_X}\bigg|_{s, n_B} \right)^{-1},\\
        &\zeta_0 = - \frac{n_B}{\kappa} \left( \frac{\partial \delta \mu_X}{\partial Y_X}\bigg|_{s, n_B} \right)^{-1} \nonumber \\
        &\quad\quad\quad \times P_1 \left( n_B \left.\frac{\partial \delta \mu_X}{\partial n_B}\right|_{s, Y_X} + s \frac{\partial \delta \mu_X}{\partial s}\bigg|_{n_B, Y_X} \right), \\
        &\de_{\Pi\Pi} = - \frac{n_B}{\kappa P_1} \left( \frac{\partial \delta \mu_X}{\partial Y_X}\bigg|_{s, n_B} \right)^{-1} \nonumber \\
        &\quad\quad\quad \times \left( n_B \left.\frac{\partial P_1}{\partial n_B}\right|_{s, Y_X} + s \frac{\partial P_1}{\partial s}\bigg|_{n_B, Y_X} \right).
    \end{align}
\end{subequations}
where $\tau_\Pi$ is the bulk relaxation time, $\zeta_0$ is the bulk viscosity, and $\de_{\Pi\Pi}$ is a second-order hydrodynamic coefficient. If we only keep the leading-order transport coefficients in $\de\mu_X$, Eq.~\eqref{eq:ISfinalform} can be simplified to
\begin{equation}
\label{IS:T0}
    u^\mu \nabla_\mu \Pi = - \frac{\Pi}{\tau_\Pi} - \frac{\zeta_0}{\tau_\Pi} \nabla_\mu u^\mu.
\end{equation}

Assuming small-amplitude baryon-density oscillations $\propto e^{-i\omega t}$ about a spatially uniform, $\beta$-equilibrated background, linear response yields
\begin{align}\label{eq:linear_response}
\delta \Pi= - \frac{\zeta_0}{1-i\tau_\Pi \omega}\,\delta(\nabla_\mu u^\mu)
= \frac{i}{\omega}\,G^{\theta}_R(\omega)\,\delta(\nabla_\mu u^\mu)\,,
\end{align}
where $G^{\theta}_R(\omega)$ is the retarded Green's function relating the bulk stress to the expansion scalar (see \cite{Yang:2023ogo, Yang:2025yoo} for details). Here $\omega$ is the local forcing angular frequency, which should be distinguished from the gravitational-wave frequency $f$.

The retarded Green's function can be decomposed into a real part (reactive, conservative response) and an imaginary part (dissipative response) \cite{Yang:2025yoo}. 
Physically, this reactive part is the additional stiffness the bulk response acquires when the composition cannot re-equilibrate within an oscillation period---equivalently, the difference between the adiabatic (frozen-composition) and equilibrium ($\beta$-equilibrated) sound speeds~\cite{Gavassino:2020kwo}. The same reactive stiffening provides the buoyancy that supports composition $g$-modes~\cite{1992ApJ...395..240R}, and for $npe$ matter it is controlled, like the bulk viscosity itself, by the density dependence of the symmetry energy~\cite{Yang:2025yoo}.
The frequency-dependent bulk viscosity then follows,
\begin{equation}
\label{eq:kubozeta}
    \zeta(\omega)=\frac{1}{\omega}\,\operatorname{Im} G^{\theta}_R(\omega)
    =\frac{1}{1+\omega^2\tau_\Pi^2}\,\zeta_0\,,
\end{equation}
which has the same functional form as the classic weak-interaction bulk-viscosity result \cite{Sawyer:1989dp}. 

In Fig.~\ref{fig:bulk-viscosity-example-plots}, we illustrate the $T$ dependence of the frequency-dependent bulk viscosity
$\zeta(\omega)$ at a representative driving frequency $\omega = 2\pi \times 1\,\mathrm{kHz}$ for both $npe$
and quark matter at $n_B = 3.5\,n_{\rm sat}$ and $n_B = 5\,n_{\rm sat}$. The red and green curves show that, for $npe$
matter, increasing either $L_{\rm sym}$ or $K_{\rm sym}$, as well as
increasing $n_B$, systematically enhances $\zeta(\omega)$ over most of the $T$ range while
preserving the characteristic single-peak structure set by the competition between the weak-interaction
relaxation time and the forcing period, consistent with previous estimates \cite{Sawyer:1989dp,Haensel:1992zz,Alford:2017rxf}. The purple curves display the corresponding results for the three-flavor
quark matter, where $\zeta(\omega)$ exhibits a qualitatively similar peak structure but is much less
sensitive to variations in $B,n_B$, with only a modest enhancement for the
largest bag constant at $n_B=5\,n_{\rm sat}$. The peak in $\zeta(\omega)$ for quark matter is shifted to significantly lower
$T$ compared to $npe$ matter, reflecting the different $T$ dependence of the underlying
strangeness-changing weak processes, consistent with standard quark-matter estimates \cite{Madsen:1992sx, Sad:2007afd, Alford:2017rxf}. Similar trends will be observed in the macroscopic dissipative tidal response in the following sections.
\section{Tidal perturbations of a neutron star}\label{sec:tidal_perturb}

In this section, we review dynamical tidal interactions of a non-rotating neutron star and describe how we calculate the tidal response function. 
The organization of this section is as follows.
We begin this section by describing the tidal response of a neutron star in a binary system in Sec.~\ref{sec:tidal-response-review} and present the equations for calculating the tidal response function in Secs.~\ref{sec:TOV-equations} and~\ref{sec:tidal-perturbations}. 
Only the basic ingredients of the procedure are presented below; for additional discussion and a detailed description, we direct the reader to~\cite{HegadeKR:2024agt,HegadeKR:2025qwj}.
\subsection{Review of tidal response theory}\label{sec:tidal-response-review}
In this section, we review the tidal response theory of a neutron star in a binary system and discuss how the tidal response coefficients affect the gravitational waveform.
Our discussion is based on~\cite{HegadeKR:2024agt,Poisson:2020vap,Ripley:2023qxo}, and we refer the reader to those references for a more detailed discussion.
In this section, we adopt the following notation.
We work with Cartesian coordinates $\left(t, x, y, z \right)$.
The indices $\left(i,j,\ldots\right)$ are used to denote spatial coordinates only.
We denote the $N^{th}$ multipole moment of body $A$ by $I^{N}_{A}$ where $N=i_1,...i_{N}$ is a multi-index of size $N$.
We denote the mass and radius of objects $A$ and $B$ by $m_{A/B}$ and $R_{A/B}$.
We occasionally denote time derivatives with an overhead dot.
We use $\stf{\cdots}$ in index lists to denote the symmetric trace-free (STF) combination of tensorial indices.
We transform from physical space to Fourier space by replacing $\partial_t \to -i \omega$.

Consider two neutron stars in a generic orbit. To leading order in a small velocity expansion, their motion can be descried in Newtonian theory in the center-of-mass frame\footnote{For a discussion of tidal interactions in post-Newtonian theory, see~\cite{Racine_2005,Vines_flanagan_2010,Vines:2011ud,Mandal:2023hqa}.}.
We model the binary as a system of two point particles, and introduce finite size corrections order by order in a multipolar expansion of each object (for a review, see Chapter 1 of~\cite{Poisson-Will}).
Restricting to linear multi-polar interactions of the objects, the center-of-mass acceleration of the binary is 
\begin{align}
\label{eq:newtonian-eob-acceleration}
    a_i
    =
    &
    -
    \frac{M}{d^2}n_i
    +
    \sum_{\ell = 2}^{\infty}
    \frac{M}{\ell!}\left(
        \frac{I_A^{\stf{L}}}{m_A}
        +
        (-1)^{\ell}
        \frac{I_B^{\stf{L}}}{m_B}
    \right)
    \partial_i\partial_{L}\frac{1}{d}
    ,
\end{align}
where $a^i\equiv \ddot{x}^i_A - \ddot{x}^i_B$ is the relative acceleration, $M$ is the total mass, $d$ is the distance between the two objects, and $n_i$ is the normal vector.
We assume the binary is evolving in a circular orbit of frequency $\omega_{\mathrm{orb}}$.

To close the system of equations, we must relate the multipolar moments of each star to the gravitational field external to it.
A common assumption is that we can model the tidal response of the neutron star via linear response theory.
We can then relate the STF multipole moments of object $A$ to the external field experienced by $A$ due to $B$, $\mathcal{E}_A^{L}$, via the tidal response function $K_{\ell}(t-t')$ through 
\begin{align}
    I_A^{\stf{L}}(t) =  -\frac{2 }{(2\ell-1)!!} R_A^{2\ell + 1}\int_{-\infty}^{\infty} K_{\ell} (t-t') \mathcal{E}_A^{L} (t') dt'\,.
    \nonumber
\end{align}
In Fourier space this integral equation becomes
\begin{align}\label{eq:tidal-response-definition-v1}
    \hat{I}^{\stf{L}}_A(\omega)
    =
    -\frac{2}{(2\ell-1)!!} R_A^{2\ell + 1} \hat{K}_{\ell}(\omega) \hat{\mathcal{E}}_A^{L}
    \,.
\end{align}
We now switch from an STF-tensor description to a spherical-harmonic description, using the relation (see, e.g. Sec. 1.5.2 of~\cite{Poisson-Will})
\begin{subequations}\label{eq:multipole-moment-driving-potential-def}
\begin{align}
    \hat{I}^{\stf{L}}_A(\omega) &\equiv \frac{4 \pi \ell!}{(2\ell + 1)!!} \sum_{m=-\ell}^{\ell} \mathscr{Y}^{* \stf{L}}_{\ell m} \hat{I}_{A,\ell m}(\omega) \,,\\
    \hat{\mathcal{E}}_A^{L} &\equiv -\frac{4\pi \ell!}{2 \ell + 1} \sum_{m=-\ell}^{\ell} d_{A,\ell m} (\omega) \mathscr{Y}^{* \stf{L}}_{\ell m}
    \,,
\end{align}
\end{subequations}
where $\mathscr{Y}^{* \stf{L}}_{\ell m}$ is the STF tensor that transforms Cartesian STF tensors into spherical harmonics, $\hat{I}_{A,\ell m}$ is the Fourier transform of the multipole moment of object $A$, and $d_{A,\ell m}$ is the tidal driving potential felt by object $A$ due to object $B$.
From here on, we drop the labels $A/B$ to simplify our notation, unless the context makes it unclear.

Using the above definitions, Eq.~\eqref{eq:tidal-response-definition-v1} can be simplified to 
\begin{align}\label{eq:tidal-response-definition}
    \hat{I}_{\ell m}(\omega)
    =
    2 R^{2\ell + 1} \hat{K}_{\ell}(\omega) d_{ \ell m}(\omega)
    \,.
\end{align}
where note that the tidal response function $\hat{K}_{\ell}(\omega)$ is, in general, a complex function that depends on the internal structure of the compact object and the frequency of the orbit.
An elementary property of $\hat{K}_{\ell}(\omega)$ follows from the fact that $K_{\ell}(t)$ is a real function,
\begin{align}
    \hat{K}_{\ell}^{*}(\omega) = \hat{K}_{\ell}(-\omega)\,,
\end{align}
and therefore, 
\begin{subequations}
\begin{align}
    \mathrm{Re}\left(\hat{K}_{\ell}(\omega)\right)
    =
    \frac{\hat{K}_{\ell}(\omega) + \hat{K}_{\ell}(-\omega)}{2} \,,\\
    \mathrm{Im}\left(\hat{K}_{\ell}(\omega)\right)
    =
    \frac{\hat{K}_{\ell}(\omega) -\hat{K}_{\ell}(-\omega)}{2i} \,.
\end{align}
\end{subequations}
The real part of the tidal response is an even function of $\omega$ and it quantifies interactions that conserve orbital energy, while the imaginary part probes dissipative interactions.
Henceforth, we will call $\mathrm{Re} \left(\hat{K}_{\ell}\right)$ the \textit{conservative tidal response function} and $\mathrm{Im} \left(\hat{K}_{\ell}\right)$ the \textit{dissipative tidal response function}.
The dominant contribution to the conservative tidal response function are the fluid oscillations inside the neutron star.
The dissipative tidal response function receives its dominant contribution from dissipative processes, such as viscosity. 
For the $npe$ EoS considered in Sec.~\ref{sec:NPE-EOS}, dissipation arises due to Urca processes, while in the quark EoS case, dissipation is due to strangeness-changing reactions. 

To understand the impact of the tidal response function on the gravitational waveform, we can expand the conservative and dissipative tidal response functions in a Taylor series about $\omega=0$\footnote{We note that this description is not strictly valid during the excitation of fluid resonances, such as $g$-modes and $f$-modes, see Sec.~\ref{sec:results} for a discussion.},
\begin{subequations}
\begin{align}
    \mathrm{Re} \left(\hat{K}_{\ell}(\omega)\right) &= k_{\ell}(0) + \omega^2 k_{\ell}^{(2)} + \mathcal{O}(\omega^4) \,,\\
    \mathrm{Im} \left(\hat{K}_{\ell}(\omega)\right) &= k_{\ell}^{(1)}(0) \; \omega + \mathcal{O}(\omega^3)\,,
\end{align}
where we have absorbed the factor of $2!$ into the coefficient $k_\ell^{(2)}$.
We can combine these two expansions to obtain
\end{subequations}
\begin{align}\label{eq:low-frequency-limit}
    \hat{K}_{\ell}(\omega) = k_{\ell}(0) + ik_{\ell}^{(1)}(0)\omega + k_{\ell}^{(2)} \omega^2 + \mathcal{O}(\omega^3) \,.
\end{align}
The constant $k_{\ell}(0)$ is the ($\ell$th) static tidal Love number, $k_{\ell}^{(1)}(0)$ is the dissipative tidal Love number, and $k_{\ell}^{(2)}$ is a dynamical tidal Love number \cite{1994MNRAS.270..611L,HegadeKR:2023glb,Steinhoff:2016rfi}. 
Henceforth, we shall restrict our attention to the dominant quadrupolar ($\ell=2$) contributions.

These different Love numbers first enter the gravitational-wave phase at different PN orders.
The static tidal Love number first appears in the gravitational waveform at $5$ PN order and has been used extensively to constrain the equilibrium nuclear EoS using gravitational-wave observations~\cite{Hinderer_2010,Chatziioannou_2020}.
The dissipative tidal Love number 
enters the gravitational waveform first at $4$ PN order, and has been used recently to inform the magnitude of dissipative effects, such as bulk- and shear-viscous dissipation~\cite{Ripley:2023lsq,Ripley:2023qxo}.
The dynamical tidal Love number $k_{\ell=2}^{(2)}$ enters the gravitational waveform first at $8$ PN order, and is determined mainly by the $f$-mode oscillations, and the viscous relaxation times that damp the $f$-mode excitation~\cite{Hinderer:2016eia,Pratten:2021pro}.
This low-frequency characterization of the tidal response function is strictly only valid during the early inspiral. 

In this paper, we shall not adopt the small frequency approximation. We calculate the full frequency-dependent tidal response function.
We also note that the gravitational waveform is affected by a compactness-weighted tidal response function that is usually called the tidal deformability~\cite{Flanagan:2007ix,Vines:2011ud,Ripley:2023lsq}.
These are related to the response function via
\begin{subequations}
\begin{align}
\label{eq:conservative_tidal_deformability}
    \Lambda(f)
    &= \frac{2 \mathrm{Re} \left(\hat{K}_{2}(f)\right)}{3 C^5} \,,\\
    \Xi(f) 
    &= 
    \frac{2 \mathrm{Im} \left(\hat{K}_{2}(f)\right)}{6 C^6 \pi f R}
    \,,
\end{align}
\end{subequations}
where $\Lambda(f)$ is the frequency-dependent conservative tidal deformability 
and $\Xi(f)$ is the frequency-dependent dissipative tidal deformability.
To calculate these frequency-dependent tidal deformabilities, we first need to calculate the frequency-dependent tidal response function $K_2(f)$, which is related to the induced multipole moment through Eq.~\eqref{eq:tidal-response-definition}.
To extract the induced multipole moment, we must understand the linear response of an initially non-rotating Tolman-Oppenheimer-Volkoff (TOV) star to an external tidal field~\cite{HegadeKR:2024agt}, which we review next.
\subsection{TOV equations}\label{sec:TOV-equations}
The background metric of a neutron star can be represented through the line element
\begin{align}\label{eq:background-metric}
    ds^2_{\mathrm{bkg}} 
    = 
    -
    e^{\nu(r)} dt^2 
    + e^{\lambda(r)} dr^2 
    + r^2 d\Omega^2 
\end{align}
in Schwarzschild-like coordinates. 
For simplicity, we describe how we find the equilibrium TOV solutions for the $npe$ EoS model here. The same technique generalizes to the quark EoS.

The Einstein equations, combined with the fluid equations, simplify to the TOV equations
\begin{subequations}\label{eq:TOV-equations}
\begin{align}
    \label{eq:lambda-equation-TOV}
    \lambda'(r) &= \frac{1- e^{\lambda(r)} + 8 \pi r^2 e^{\lambda(r)} \varepsilon}{r} \,,\\
    \label{eq:nu-equation-TOV}
    \nu'(r) &= \frac{-1+e^{\lambda }+8 e^{\lambda } \pi  r^2 P}{r} \,,\\
    \label{eq:p-equation-TOV}
    P'(r) &= -\frac{(\varepsilon+P) \left(-1+e^{\lambda }+8 e^{\lambda } \pi  r^2 P\right)}{2 r }
    \,.
\end{align}
\end{subequations}
No equation is needed for the evolution of the electron fraction because the background is in $\beta$-equilibrium, which allows us to write $Y_X = Y_X^{\mathrm{eq}}(T, \rho)$, where $\rho=m_B\,n_B$ is the mass density.
Using the variables $P\left(T, \rho,Y_X\right)$ and $\varepsilon(T, \rho,Y_X)$, we can rewrite Eq.~\eqref{eq:p-equation-TOV} as
\begin{align}
     \label{eq:p-equation-TOV-1}
    \frac{d\rho}{dr} &\left[ \left. \frac{\partial P }{\partial \rho} \right|_{\mathrm{eq},T,Y_X} 
    + 
    \left.
    \frac{\partial P }{\partial Y_X}
    \right|_{\mathrm{eq},T,\rho}
    \frac{\partial Y_X^{\mathrm{eq}} }{\partial \rho} \right]
    \nonumber\\
    =& 
    -\frac{(\varepsilon + P) \left(-1+e^{\lambda }+8 e^{\lambda } \pi  r^2 P\right)}{2 r }
    \nonumber\\
    &-
    \frac{dT}{dr}
    \left[
    \left.
    \frac{\partial P }{\partial T}
    \right|_{\mathrm{eq},\rho,Y_X}
    + 
    \left.
    \frac{\partial P }{\partial Y_X}
    \right|_{\mathrm{eq},T,\rho}
    \frac{\partial Y_X^{\mathrm{eq}} }{\partial T} \right]
    \,.
\end{align}
Given a temperature profile $T(r)$, we can integrate the above equations to obtain the TOV solution.
Two common choices for the profile are~\cite{Gittins:2024oeh}: i) a constant temperature profile $T(r) = T_0$ and ii) a redshifted temperature profile $T(r) e^{\nu/2} = T_0$ where, $T_0$ is a constant in both cases.
In this work, for simplicity, we assume that the star is at a constant temperature. 
\subsection{Tidal perturbations}\label{sec:tidal-perturbations}
The tidal field of the external universe (which, in this case, is created by the binary companion) causes a perturbation to the spherically-symmetric metric of a TOV solution.
These perturbations are decomposed in terms of Fourier modes and spherical harmonic components in the Regge-Wheeler gauge (see Sec. IV of~\cite{HegadeKR:2024agt} for more details).
For simplicity, we only consider non-radial, polar perturbations of the background metric in Eq.~\eqref{eq:background-metric}, since these generate leading-order (electric-type) effects in the gravitational waveform.
Decomposing the linear perturbations to the metric components into spherical harmonics $Y_{\ell m}$, we obtain the perturbed line element as
\begin{align}\label{eq:metric-polar-pert}
    ds^2 &= -e^{\nu(r)} \left(1 - 2 H(r) e^{-i\omega t}  r^{\ell} Y_{\ell m}\right) dt^2
    \nonumber\\
    &- 
    2 i H_1(r) e^{-i\omega t}  r^{\ell} Y_{\ell m} dt dr
    \nonumber\\
    &+
    e^{\lambda(r)}
    \left(1 + 2 H_2(r) e^{-i\omega t}  r^{\ell} Y_{\ell m }\right) dr^2
    \nonumber\\
    &+
    r^2 \left(
    1 - K(r) e^{-i\omega t} r^{\ell} Y_{\ell m }
    \right)
    d \Omega^2
    \,.
\end{align}

Let us now discuss the perturbations to the fluid.
We use $\delta$ and $\Delta$ to denote the Eulerian and Lagrangian perturbations of the fluid variables; these are related by 
\begin{align}
    \Delta = \delta  + \mathcal{L}_{\xi}
\end{align}
where $\mathcal{L}_{\xi}$ denotes the Lie derivative along the displacement vector $\xi^{\mu}$ of the fluid.
We parameterize non-zero components of the Lagrangian displacement vector as 
\begin{subequations}
\begin{align}\label{eq:lagrangian-displacement-vectors}
    \xi^r &= W(r) e^{-i\omega t} r^{\ell - 1} e^{-\lambda/2} Y_{\ell m} \,,\\
    \xi_A &= - V(r) e^{-i\omega t} r^{\ell } E_{A}^{\ell m} \,,
\end{align}
\end{subequations}
where the index $A\in (\theta,\phi)$ and $E_{A}^{\ell m}$ are vector spherical harmonics. The functions $(W(r),V(r))$ will be determined by solving the Einstein-fluid equations.
The Eulerian perturbation of the four-velocity can be determined from the Lagrangian displacement using~\cite{FN-book} 
\begin{align}
    \Delta u^{\alpha}
    =
    \frac{1}{2}u^{\alpha}u^{\beta}u^{\gamma}\Delta g_{\beta\gamma} \,.
\end{align}
We can simplify the above equation using Eq.~\eqref{eq:lagrangian-displacement-vectors} to find
\begin{subequations}
\begin{align}\label{eq:pertubed-four-veolocity-vectors}
    \delta u^t 
    &=  
    e^{-\frac{\nu }{2}} r^{\ell } H e^{-i \omega t} Y_{\ell m }
    \,,\\
    \delta u^r 
    &= 
    -i \omega e^{-(\lambda+\nu)/2} r^{\ell-1 } W e^{-i \omega t} Y_{\ell m } 
    \,,\\
    \delta u_{A} 
    &= 
    i \omega e^{-\nu/2} r^{\ell} V e^{-i \omega t} E^{\ell m}_A 
    \,.
\end{align}
\end{subequations}

Before proceeding, we need a relation between the Lagrangian perturbation of the pressure and the change in mass density.
We parameterize this as
\begin{subequations}
\begin{align}
    \frac{\Delta \rho}{\rho} 
    &= 
    -
    \frac{1}{2} \left(g^{\alpha \beta} + u^{\alpha} u^{\beta} \right) \Delta g_{\alpha \beta}
    \,,\\
    \frac{\Delta P}{P}
    &= \gamma_{1} \frac{\Delta \rho}{\rho} + \gamma_{2} \Delta Y_X\,,
\end{align}
\end{subequations}
where
\begin{subequations}
\begin{align}
    \gamma_{1} &= \frac{\rho}{P}\left.\frac{\partial P}{\partial \rho}\right|_{s,Y_X}\,,\\
    \gamma_{2} &=  \frac{1}{P}\left.\frac{\partial P}{\partial Y_X}\right|_{s,\rho} \,.
\end{align}
\end{subequations}
The coefficient $\gamma_{1}$ is called the adiabatic index of the star.
We describe the calculations of these thermodynamic derivatives in Appendix~\ref{appendix:derivatives}.

At this point, the unknown functions that we need to determine from the field equations are $\left(H, W, V, H_1,H_2,K, \delta Y_X\right)$.
To reduce the number of free variables further, we can use the linearized Einstein equations.
Let us use the variables $E^{\alpha \beta}$ and $E^{a}$ to denote the linearized Einstein equations and the stress-energy conservation equations respectively,
\begin{subequations}
\begin{align}
    E_{\alpha \beta} &\equiv \delta G_{\alpha \beta} - 8 \pi \delta T_{\alpha \beta} \,,\\
    E^{\alpha} &\equiv \delta \left(\nabla_{\beta} T^{\beta \alpha} \right) \,.
\end{align}
\end{subequations}
We first find a relation for $H_2$ by using the $(\theta,\phi)$ component of the linearized Einstein equation 
\begin{align}\label{eq:H2-elim}
    H_2 &= H\,.
\end{align}
Next, we use Eq.~\eqref{eq:H2-elim} to eliminate $H_2$ in the $(r,\theta)$ and $(t,\theta)$ components of $E_{\alpha \beta}$ and obtain expressions for $H_1$ and $H_1'$, see Eq. (30) of~\cite{HegadeKR:2024agt}.
We can eliminate $K$ and $K'$ by using the $(r,r)$ component and the $(t,r)$ component of the field equations.
Finally, one can eliminate $\delta Y_X$ using the evolution equation for the species fraction, leaving us with a set of master variables
\begin{align}
    \vec{\boldsymbol{Y}} \equiv \left(H,W,V,H'\right) \,.
\end{align}
The master variables satisfy a linear system of equations, which can be schematically written as
\begin{align}\label{eq:master-equation-matrix}
    \Vec{Y}' &= \mathbf{A} \Vec{Y}\,,
\end{align}
where, the matrix $\mathbf{A}$ is a complex matrix that depends only on the background TOV quantities.
We provide explicit expressions for these functions in a supplementary \texttt{Mathematica} notebook available upon release. 
We note that the set of master equations we have described above is not unique and there are a number of different but equivalent master equations one can solve, see~\cite{1983ApJS...53...73L,Lindblom_1997,FN-book} for other approaches.

The master equations have to be combined with suitable boundary conditions at the center and the surface of the star to obtain the tidal response function of the object.
At the center of the star, we demand that the functions $(H,W,V)$ be smooth functions of the radial coordinate $r$.
At the surface of the star, we demand that the fluid variables $(W,V)$ be finite and differentiable, and that $H(r)$ match, in a continuous and differentiable manner, the external metric function $H_{\mathrm{ext}}$ that describes the tidal deformations of the star~\cite{HegadeKR:2024agt}.
The explicit form of $H_{\mathrm{ext}}$ is provided in the supplementary \texttt{Mathematica} notebook. For a detailed derivation, see~\cite{HegadeKR:2024agt,Pitre:2023xsr}.
Matching the internal solution to the external solution at the surface of the star determines the frequency-dependent tidal response function $\hat{K}_2(f)$.
\section{Results}\label{sec:results}
In this section, we discuss how the frequency-dependent tidal response function $\hat{K}_2(f)$ depends on the EoS described in Secs.~\ref{sec:NPE-EOS} and~\ref{sec:quark-EOS}.
Let us first describe how we explore the multi-dimensional parameter space of the $npe$ EoS and the quark EoS. 
We explore how different physical quantities depend on five parameters of the $npe$ model, given by
\begin{align*}
    \theta_{\mathrm{NPE}}
    &=
    \left(T, K_0, S_{\mathrm{sym}}, L_{\mathrm{sym}}, K_{\mathrm{sym}} \right)\,.
\end{align*}
Here we require that all choices of these parameters must allow for mass-radius curves that pass known constraints. 
As such, while $K_{\mathrm{sym}}$ is entirely unconstrained by experiments, we use tighter ranges that are largely constrained by the range of $ L_{\mathrm{sym}}$. 
We choose a reasonable range to present our results: 
\begin{subequations}\label{eq:parameter-range-NPE}
\begin{align}
    T &\in \left[0.1, 3\right] 
    \mathrm{MeV}\,,\\
    K_0 &\in \left[220, 260\right] 
    \mathrm{MeV}\,, \\
    S_{\mathrm{sym}} &\in \left[29, 34\right] \mathrm{MeV}\,, \\
    L_{\mathrm{sym}} &\in \left[40, 60\right] 
    \mathrm{MeV}\,,\\
    K_{\mathrm{sym}} &\in \left[0, 200\right] 
    \mathrm{MeV}\,.
\end{align}
\end{subequations}
These values are consistent with experimental and theoretical constraints \cite{Colo:2013yta, Todd-Rutel:2005yzo, Colo:2004mj, Agrawal:2003xb, Khan:2012ps, Stone:2014wza, Li:2019xxz, Xie:2020tdo, Dutra:2012mb, Dutra:2014qga, Tagami:2022xvs, Kortelainen:2010hv, Steiner:2004fi, Reed:2021nqk, Brown:2000pd, Typel:2001lcw, Xu:2020fdc, Zhang:2013wna, Drischler:2020hwi, Tews:2024owl, PREX:2021umo, CREX:2022kgg, Lattimer:2023rpe, Furnstahl:2001un, Danielewicz:2013upa, Zhang:2022bni, Reinhard:2022inh, Lattimer:2023rpe, MUSES:2023hyz}. 
We also consider a complementary range of parameters for $L_{\mathrm{sym}}$ and $K_{\mathrm{sym}}$ motivated by PREX‑II (see ~\cite{Lattimer:2023rpe} for a detailed discussion, also including CREX), 
\begin{subequations}\label{eq:parameters-PREX-II}
\begin{align}
    L_{\mathrm{sym}} &\in \left[100, 120\right] 
    \mathrm{MeV}\,,\\
    K_{\mathrm{sym}} &\in \left[-50 , 0\right] 
    \mathrm{MeV}
    \,.
\end{align}
\end{subequations}
The negative values of $K_{\mathrm{sym}}$ is used in this case to pass the mass-radius constraints from NICER~\cite{Miller:2019cac, Riley:2019yda, Miller:2021qha, Riley:2021pdl} and LIGO observations~\cite{LIGOScientific:2017vwq, LIGOScientific:2018hze, LIGOScientific:2018cki}. We note that our $K_{\rm sym}$ range in Eqs.~\eqref{eq:parameter-range-NPE} is restricted to non-negative values, which is atypical for constraints from nuclear-physics experiments. This choice is dictated by the Padé-like ansatz of Eq.~\eqref{eq:sym_matter}: for the coefficients $(a_i,b_i)$ used here, sufficiently negative $K_{\rm sym}$ leads to unphysical behavior of the energy density at high baryon density (e.g.\ $\varepsilon < 0$), so we restrict to $K_{\rm sym} \geq 0$ for the fiducial parameter scan. The complementary PREX-II set in Eq.~\eqref{eq:parameters-PREX-II} uses the more conventional $K_{\rm sym} \in [-50,0]\,\mathrm{MeV}$, which is accessible because the larger $L_{\rm sym}$ stiffens the EoS at high densities and prevents the pathology. Since the tidal response is weighted toward densities of a few $n_{\rm sat}$, where the high-density pathology has little impact, the qualitative conclusions of our results are not sensitive to this restriction; see also the recent analysis of Ref.~\cite{Harris:2025ncu}, which finds an enhancement of the peak bulk viscosity at both ends of the $K_{\rm sym}$ range.

Exploring the $5$-dimensional parameter space for $\thetanpe$ in detail is computationally expensive. 
Instead, we explore each direction one at a time, starting from two reference sets of parameters that pass known mass-radius, static tidal deformability, and nuclear saturation constraints: a fiducial set and a large‑$L_{\mathrm{sym}}$ set motivated by PREX‑II,
\begin{align}\label{eq:theta-canonical}
    \thetanpe^{\mathrm{fid}}
    &=
    \left(0.1 , 250, 32, 50, 0 \right) \mathrm{MeV}\,,\\
    \thetanpe^{\mathrm{PREX-II}}
    &=
    \left(0.1 , 250, 32, 120, -50 \right) \mathrm{MeV}
    \,.
\end{align}
This allows us to understand how macroscopic quantities, such as the mass, radius, and the tidal response function, depend on the microscopic parameters.

The quark model is described by three parameters
\begin{align}
    \thetaquark = \left(T , m_{s} , B \right)\,.
\end{align}
We fix the mass of the strange quark to a common choice $m_{s} = 98 \;\mathrm{MeV}$ \cite{Alcock:1986hz, Haensel:1986qb}\footnote{We note that past works have shown that quark mass can also affect the bulk viscosity \cite{Hernandez:2024rxi}} and vary the temperature and the bag constant in the range
\begin{subequations}\label{eq:quark-parameter-range}
\begin{align}
    T &\in \left[ 10^{-5}, 10^{-3}\right] \mathrm{MeV} \,,\\
    B &\in \left[60, 70\right]\, \frac{\mathrm{MeV}}{\mathrm{fm}^3} \,.
\end{align}
\end{subequations}
Note that the temperature dependence in the EoS only enters through the reaction rates.

The outline of the rest of this section is as follows. In Sec.~\ref{sec:MR} we present the mass-radius ($M$--$R$) curves. 
Section~\ref{sec:Tidal-response-func} describes our results for the conservative tidal response function and discusses how the microphysics and the frequency $f$ induce variations in $\mathrm{Re}\left[ \hat{K}_2(f) \right]$.
Finally, we discuss the frequency-dependent dissipative tidal response function in Sec.~\ref{sec:dissipative-tidal-response} and the low-frequency $g$-mode resonances in Sec.~\ref{sec:g-mode-resonances}.
\subsection{Mass-radius curves}\label{sec:MR}
\begin{figure*}[t!]
    \centering
    \includegraphics[width = 0.95 \columnwidth ]{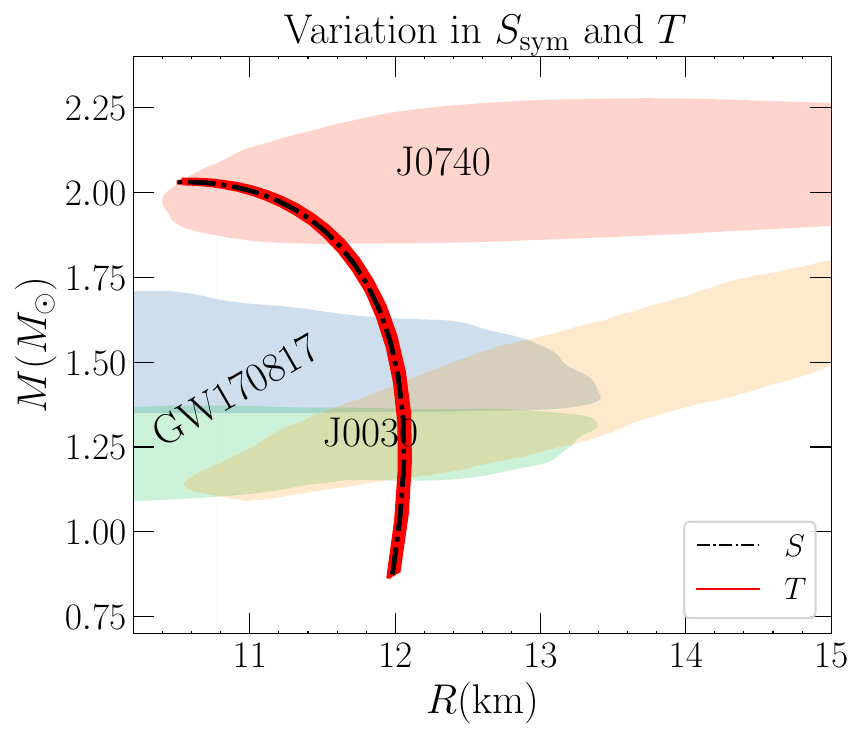}
    \includegraphics[width = 0.95 \columnwidth ]{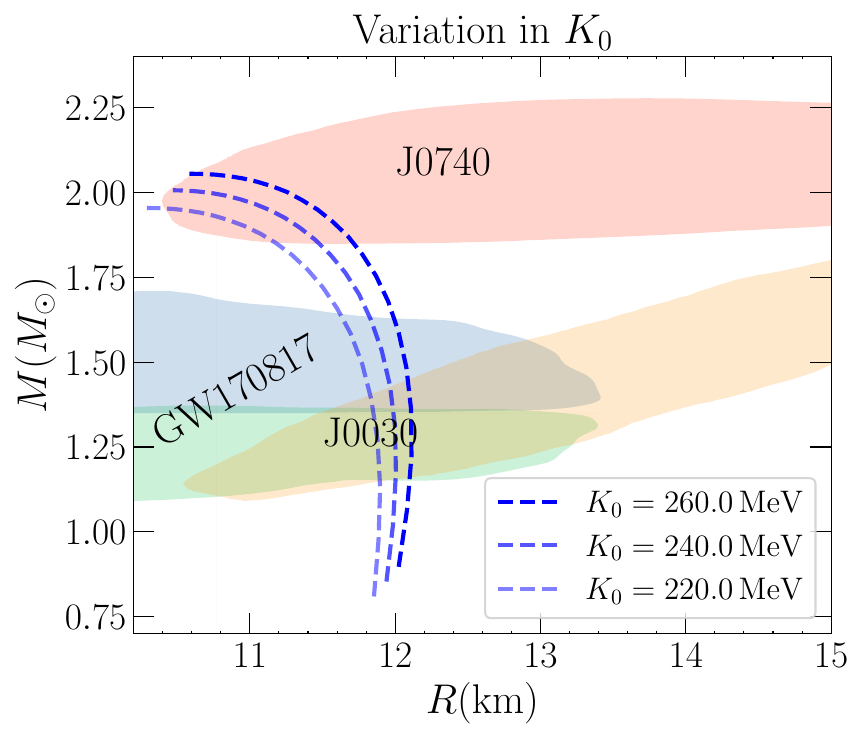}
    \includegraphics[width = 0.95 \columnwidth ]{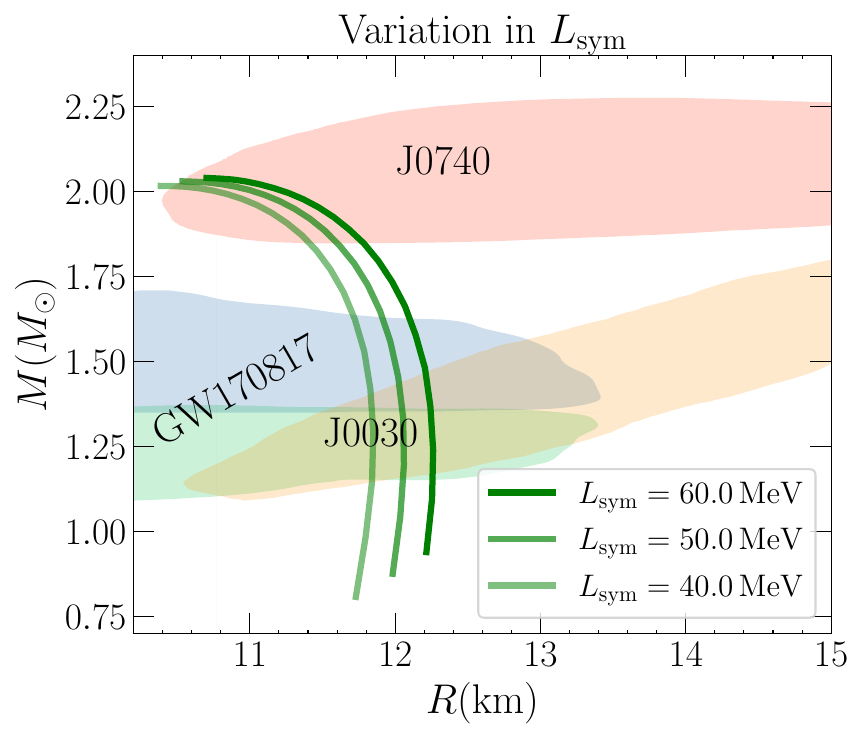}
    \includegraphics[width = 0.95 \columnwidth ]{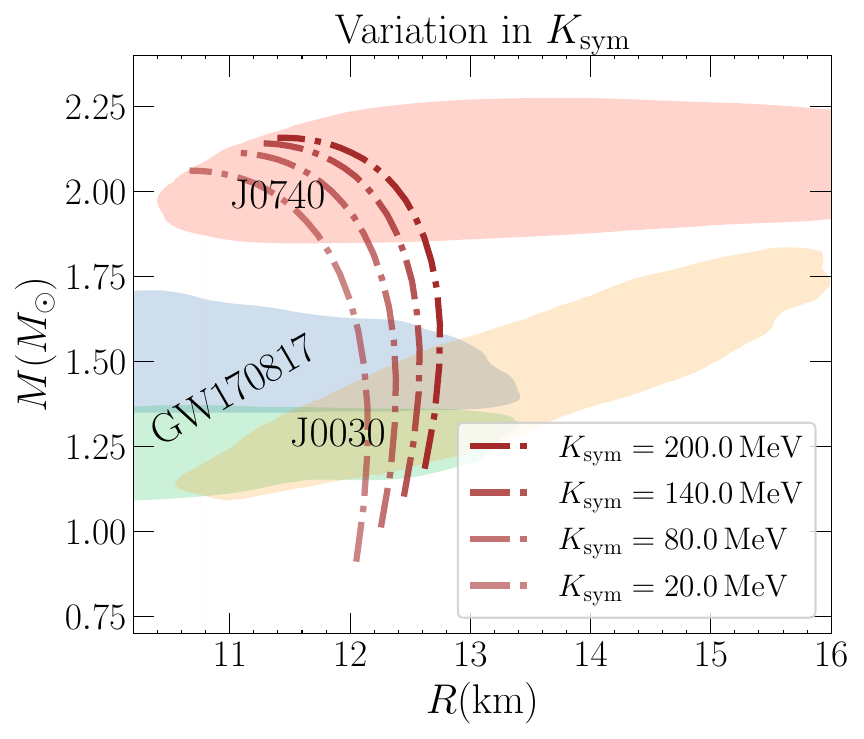}
    \caption{
    Variation of the mass--radius curves with respect to different components of $\thetanpe$,  while the remaining parameters are fixed to their fiducial values in Eq.~\eqref{eq:theta-canonical}. 
    In each panel, lighter colored curves indicate lower values of the parameter of interest. 
    The $M$--$R$ constraints from NICER~\cite{Miller:2019cac, Riley:2019yda, Miller:2021qha, Riley:2021pdl} (red--orange contours labeled with J0740 and J0030) and LIGO~\cite{LIGOScientific:2017vwq, LIGOScientific:2018hze, LIGOScientific:2018cki} (blue--green contours labeled GW170817) are also shown for comparison.
    The top, left panel explores the dependence of the $M$--$R$ curve with respect to the symmetry energy $S_{\mathrm{sym}}$ (black dash-dotted band) and the temperature $T$ (red band). Observe that the $M$--$R$ curve is almost insensitive to these parameters.
    The variation of the mass-radius curves with respect to the incompressibility $K_0$, the slope $L_{\mathrm{sym}}$ and the curvature $K_{\mathrm{sym}}$  of the symmetry energy are presented in the top right, bottom left and bottom right panel, respectively. 
    For the range considered in Eq.~\eqref{eq:parameter-range-NPE}, we see that the mass-radius curve varies significantly with respect to $K_{\mathrm{sym}}$ (bottom right) across all central densities, while the variation with respect to $L_{\mathrm{sym}}$ (bottom left) is more significant at lower central densities.
    Because the scanned ranges of $L_{\mathrm{sym}}$ and $K_{\mathrm{sym}}$ differ substantially in width, the larger absolute band for $K_{\mathrm{sym}}$ reflects its broader allowed range rather than a larger per-MeV sensitivity; per unit change, the $M$--$R$ relation is in fact more sensitive to $L_{\mathrm{sym}}$ (see Sec.~\ref{sec:Tidal-response-func}).
    }
    \label{fig:MR_curves_npe}
\end{figure*}

Let us first concentrate on $npe$ matter.
We remind the reader that the parameter ranges that we use are presented in Eqs.~(\ref{eq:parameter-range-NPE}-\ref{eq:parameters-PREX-II}).
Our results for the $M$--$R$ curves are shown in Fig.~\ref{fig:MR_curves_npe}.
\begin{figure*}
    \centering
    \includegraphics[width=0.45\linewidth]{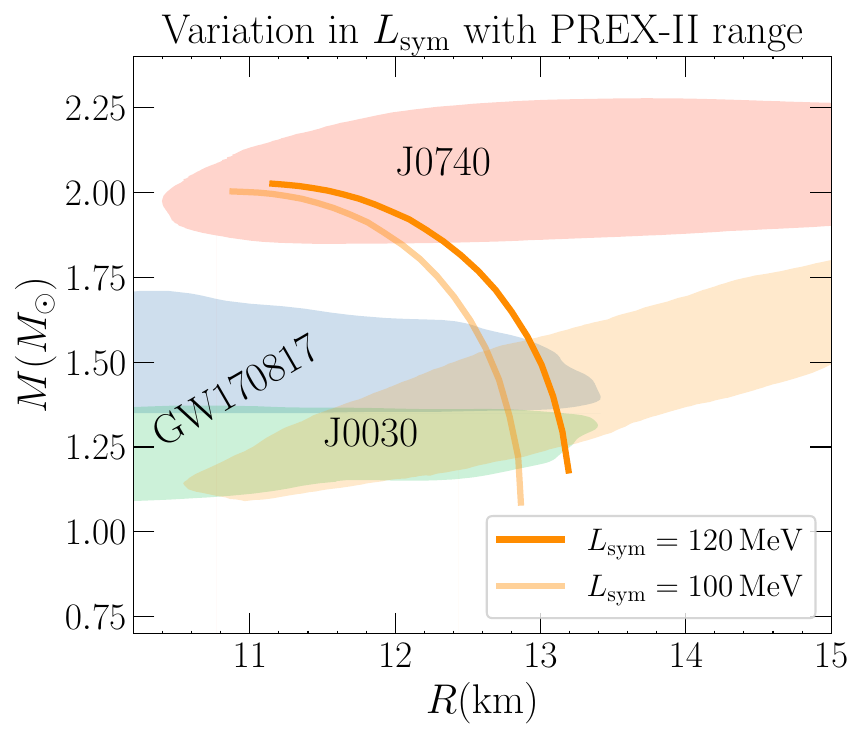}
    \includegraphics[width=0.45\linewidth]{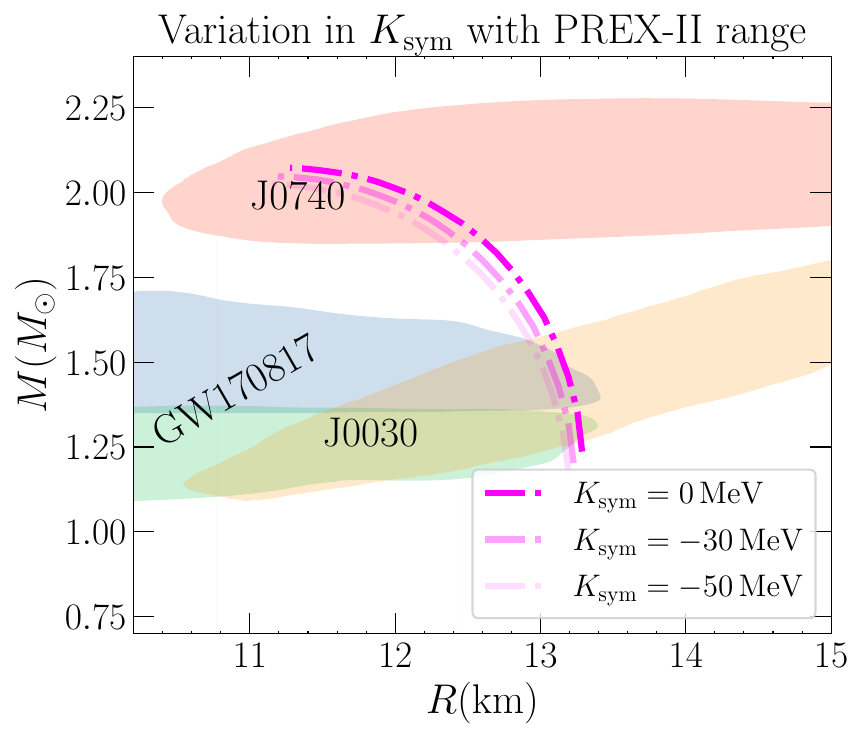}
    \caption{
    Variation of the $M$--$R$ curves with  $L_{\mathrm{sym}}$ and $K_{\mathrm{sym}}$ for parameter values for the large‑$L_{\mathrm{sym}}$ (PREX‑II‑like) range. Lighter colored curves indicate lower values of the parameter of interest. Observe that the large value of $L_{\rm sym}$ in the PREX-II measurement pushes the curves to larger radii as compared to Fig.~\ref{fig:MR_curves_npe}. The general trends of the M--R curves with $K_{\rm sym}$ and $L_{\rm sym}$, however, remain.}
    \label{fig:MR_Curves_PREX_II_range}
\end{figure*}
\begin{figure}[h!]
    \centering
    \includegraphics[width=0.95\columnwidth]{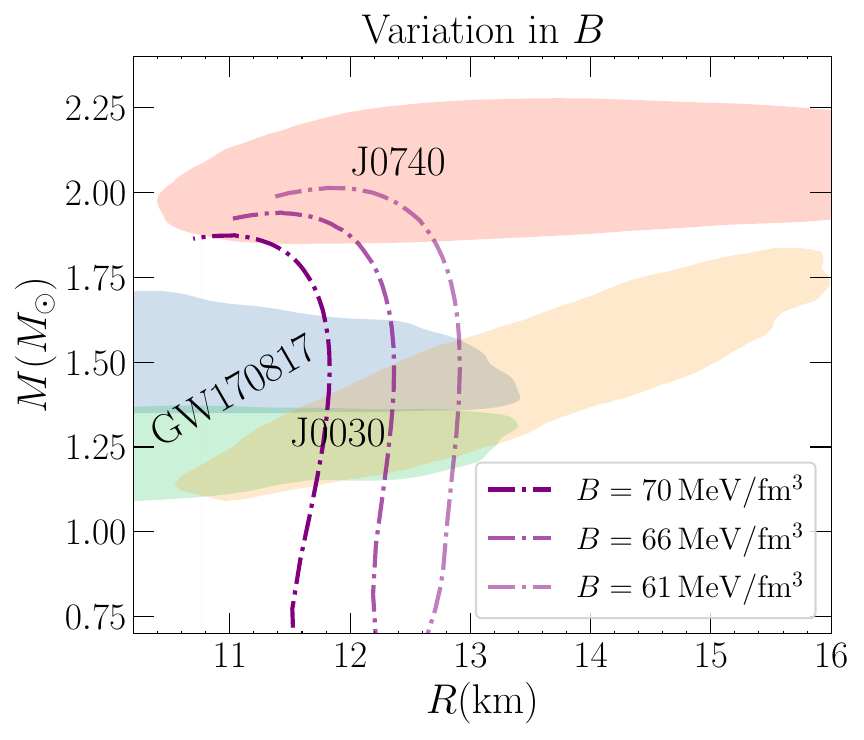}
    \caption{
    Variation of the $M$--$R$ curve of the quark EoS model with respect to the bag constant. Lighter colored curves indicate lower values of the bag constant. Observe that increasing the bag constant decreases the stiffness of the EoS, leading to a smaller mass and radius at a given number density.}
    \label{fig:quark_MR_curve}
\end{figure}
The top, left panel explores the variation of the $M$--$R$ curves with $(S_{\mathrm{sym}},T)$. 
Observe that the variation is quite minimal for the parameter range we considered. The small variation with $S_{\mathrm{sym}}$ is due to its tight experimental constraints. The small variation with $T$ is due to the weak dependence of the EoS on temperatures in the range relevant to isolated (or nearly-isolated) neutron stars, as also noted in~\cite{Gittins:2024oeh}.

The rest of the panels of Fig.~\ref{fig:MR_curves_npe} present the variation of the $M$--$R$ curves with respect to $K_0$ (top right), $L_{\mathrm{sym}}$ (bottom left) and $K_{\mathrm{sym}}$ (bottom right).
The trends shown in these panels can be understood in terms of how rapidly $\varepsilon$ increases with $n_B$ for our $npe$ EoS. 
In other words, the faster $\varepsilon(n_B)$ increases, the harder it is to compress the nuclear matter to reach $n_B$, which leads to higher values of both mass and radius. 
As one can see from Eqs.~\eqref{eq:Esym_quad}, ~\eqref{eq:sym_matter} and~\eqref{eq:model}, $\varepsilon$ increases more steeply with $n_B$ for larger values of $K_0$, $L_{\mathrm{sym}}$, and $K_{\mathrm{sym}}$. More specifically, $K_0$ controls the increase of the energy density of symmetric nuclear matter, while $L_{\mathrm{sym}}$ and $K_{\mathrm{sym}}$ contribute to the increase of the energy density of asymmetric nuclear matter as density increases. As Fig.~\ref{fig:MR_curves_npe} shows, both mass and radius increase with increasing $K_0$, $L_{\mathrm{sym}}$, and $K_{\mathrm{sym}}$.

In Fig.~\ref{fig:MR_Curves_PREX_II_range}, we also explore the variation of $L_{\mathrm{sym}}$, and $K_{\mathrm{sym}}$ for the PREX‑II‑like range. While PREX-II only constrains $L_{\mathrm{sym}}$, we must use negative values of $K_{\mathrm{sym}}$ to compensate the very stiff low $n_B$ EoS in order to reproduce constraints from GW170817. 
The same scaling with symmetry energy coefficients that we noticed in Fig.~\ref{fig:MR_curves_npe} is also seen when we pick the PREX‑II‑like parameters in  Fig.~\ref{fig:MR_Curves_PREX_II_range}. 
However, the radii found for the same values of mass are larger for PREX-II-like parameters than for the ones used in Fig.~\ref{fig:MR_curves_npe}, due to the larger value of $L_{\mathrm{sym}}$. 

In Fig.~\ref{fig:quark_MR_curve}, we show similar results for the quark model while varying $B$ (recall that for the quark model we assume that the EoS does not very strongly with $T$ such that we set $T=0$). The $M$--$R$ curves that we find are similar to those from previous calculations within the MIT bag model \cite{Alcock:1986hz, Haensel:1986qb}. 
From Eq.~\eqref{eq:quark-energy-and-number-density}, we see that, at fixed number density,  increasing the bag constant leads to an increase in energy density while simultaneously decreasing the pressure. These changes effectively decrease the stiffness of the EoS, leading to lower values of mass and radius \cite{Farhi:1984qu}, as shown in the figure.

\subsection{Conservative tidal response function}\label{sec:Tidal-response-func}

\begin{figure*}[thp!]
    \centering
    \includegraphics[width = 0.95 \columnwidth ]{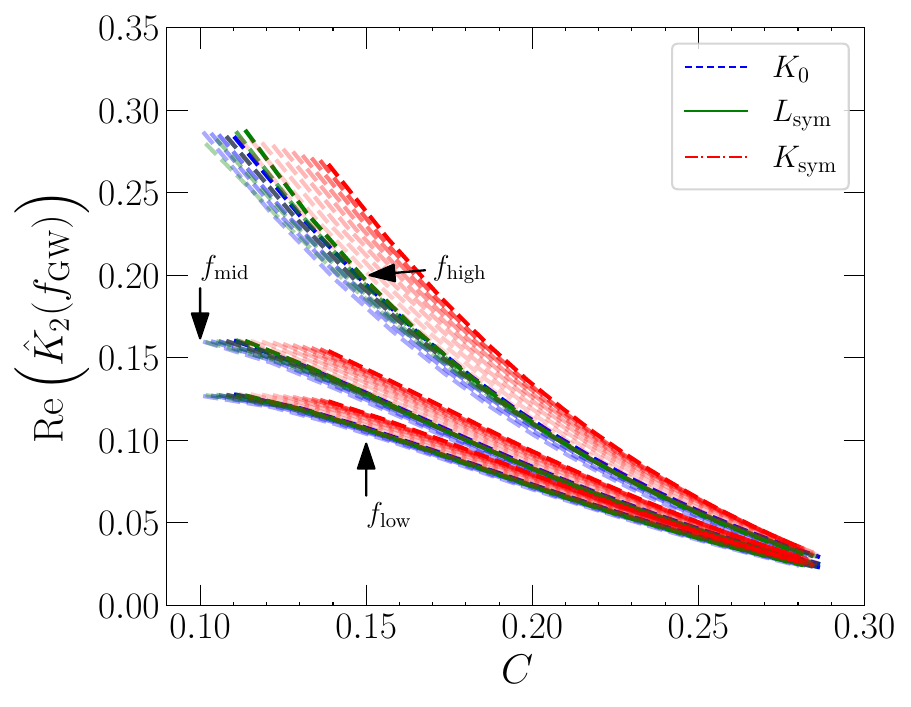}
    \includegraphics[width = 0.95 \columnwidth ]{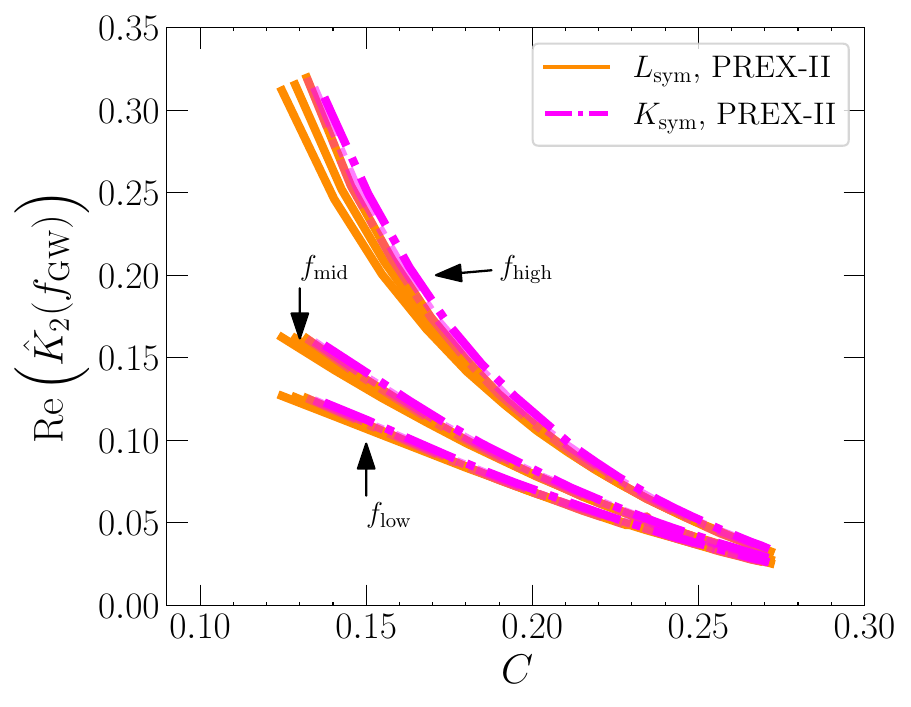}
    \includegraphics[width = 0.95 \columnwidth ]{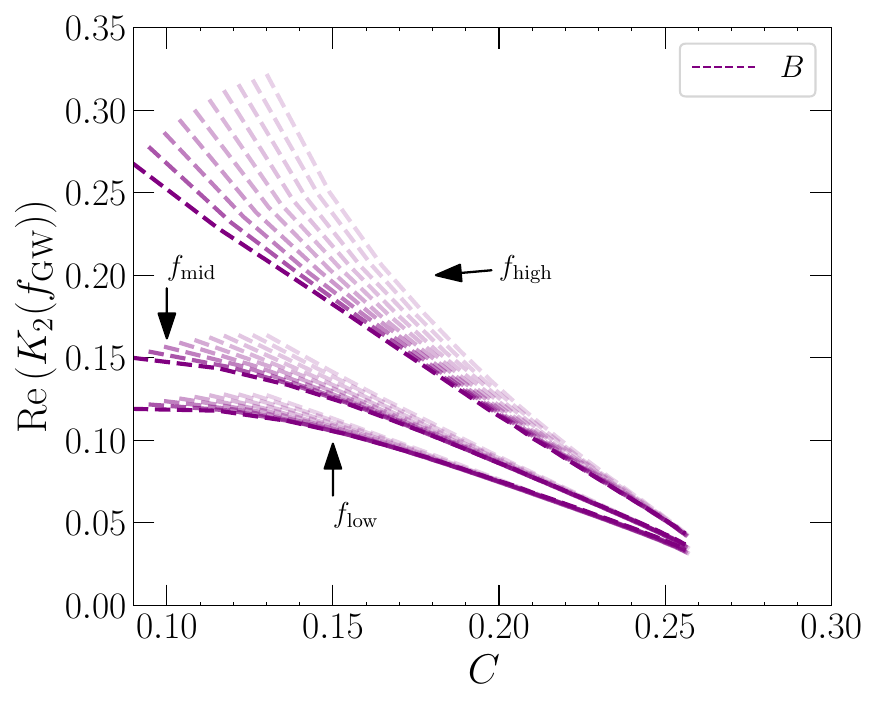}
    \caption{
     Variation of the conservative tidal response function for the $npe$ EoS, using fiducial parameter values from Eq.~\eqref{eq:parameter-range-NPE} (left-panel, blue, green, red contours) and from Eq.~\eqref{eq:parameters-PREX-II} (right-panel, orange and magenta contours), as well as for the quark EoS (bottom-panel, purple contours).
     We note that the tidal response function increases with frequency due to the excitation of the $f$-modes inside the star---this increase is more precisely quantified in Fig.~\ref{fig:conservative_tidal_response_ratio}.
    }
    \label{fig:conservative_tidal_response}
\end{figure*}

We now present results for the conservative tidal response function $\mathrm{Re}\left[ \hat{K}_2(\omega) \right]$.
This function is only sensitive to the EoS and the gravitational-wave frequency of the binary. For a binary in a quasi-circular orbit, the frequency $\omega/(2 \pi)$ is equal to the gravitational-wave frequency~\cite{Poisson-Will}. 
In the previous section, we noticed that for the $npe$ EoS, the $M$--$R$ curves are insensitive to the variations in the $(S_{\mathrm{sym}},T)$ plane (see Fig.~\ref{fig:MR_curves_npe}). 
While this is not shown here, we also found that variations in $(S_{\mathrm{sym}},T)$ do not significantly affect the conservative tidal response function. Thus, in this section, we do not consider these variations, and focus on the remaining parameters.

We study the frequency dependence on the conservative tidal response function. 
To simplify our presentation, we choose three different values for the gravitational-wave 
frequency, 
\begin{align}
    \left(f_{\mathrm{low}}, f_{\mathrm{mid}}, f_{\mathrm{high}} \right)   = \left(400 , 800, 1200 \right) \mathrm{Hz}
    \,.
\end{align}
The $f_{\mathrm{low}}$ value is the characteristic frequency around which tidal effects begin to build up in current detectors.
The $f_{\mathrm{mid/high}}$ values are used to understand how dynamical frequency-dependent effects change the tidal response function.
For current gravitational wave detectors, dynamical tidal effects are relatively weak and do not contribute significantly to the waveform.
However, starting from the fifth observing run (2030)~\cite{Pratten:2021pro,Abac_2024}, these effects will become increasingly important and ignoring them will cause a strong bias in the inference of the EoS~\cite{Pratten:2021pro}.

Figure~\ref{fig:conservative_tidal_response} presents results for the variation of the conservative tidal response function with respect to $(K_0, L_{\mathrm{sym}}, K_{\mathrm{sym}})$ for the $npe$ EoS (top row, with the upper-right panel corresponding to the PREX‑II‑like range), and with respect to the bag constant $B$ for the quark EoS (bottom row);  all are shown as a function of the stellar compactness $C=M/R$.
For the $npe$ EoS, at a given value of frequency, we see that the largest variation is due to changes in $K_{\mathrm{sym}}$ (shown as red dashed lines in the top-left panel), because $K_{\rm sym}$ remains largely unconstrained by current experiments, allowing for a wide range of variation ($\Delta K_{\rm sym} = 200\,\mathrm{MeV}$) in our parameter scan.
However, this wide absolute band may be somewhat misleading when comparing the intrinsic sensitivity to different symmetry-energy coefficients. 

To illustrate this, we focus on a fiducial $1.4\,M_\odot$ star ($C \approx 0.17$ for the fiducial parameter set) and evaluate $\mathrm{Re}[\hat{K}_2]$ at $f_{\rm low}$.
The total variation over the $K_{\rm sym}$ scan ($0$--$200\,\mathrm{MeV}$) is $\Delta\mathrm{Re}[\hat{K}_2] \approx 0.03$, while $L_{\rm sym}$ ($40$--$60\,\mathrm{MeV}$) produces $\Delta\mathrm{Re}[\hat{K}_2] \approx 0.014$.
Yet, per unit change in the parameter, the tidal response is roughly $4$--$5$ times more sensitive to $L_{\rm sym}$  than to $K_{\rm sym}$.
For the PREX-II parameter range, this hierarchy becomes even more apparent: the $L_{\rm sym}$ variation ($100$--$120\,\mathrm{MeV}$, only a $20\%$ change) produces a total shift $\Delta\mathrm{Re}[\hat{K}_2] \approx 0.02$ at $f_{\rm low}$ for a $1.4\,M_\odot$ star, which exceeds the variation from $K_{\rm sym}$ ($-50$ to $0\,\mathrm{MeV}$, $\Delta\mathrm{Re}[\hat{K}_2] \approx 0.015$) over a range which is $2.5$ times wider.
The per-MeV sensitivity of $L_{\rm sym}$ in the PREX-II regime  exceeds that of $K_{\rm sym}$ by roughly a factor of $3$.
A similar pattern holds for the stellar radius: at $1.4\,M_\odot$ in the fiducial set, $\Delta R/\Delta L_{\rm sym} \approx 0.05\,\mathrm{km/MeV}$ versus $\Delta R/\Delta K_{\rm sym} \approx 0.014\,\mathrm{km/MeV}$.
Thus, while $K_{\rm sym}$ produces the largest total band due to its wide experimental uncertainty, $L_{\rm sym}$ is intrinsically the more sensitive parameter per unit change in both the fiducial and PREX-II regimes.
Since both parameters are currently insufficiently constrained---$K_{\rm sym}$ by experiment and $L_{\rm sym}$ by the tension between PREX-II and CREX---a joint statistical analysis properly accounting for the allowed parameter volumes of each coefficient is needed to definitively rank their relative importance.

For the quark model (right-panel, purple contours), the only parameter we vary is the bag constant. 
We find that conservative tidal response of the quark model is strongly dependent on $B$ and, in fact, shows a much stronger sensitivity than what is seen with the combined changes of all the symmetry energy coefficients from the $npe$ EoS.
Additionally, we see that, similar to the $npe$ EoS, the tidal response function of the quark model also increases steeply as we increase the frequency. 

The general behavior we expect from the conservative tidal deformability as the frequency increases is~\cite{Steinhoff:2016rfi,Flanagan:2007ix}
\begin{align}
    \mathrm{Re} \left(\hat{K}_2(\omega)\right) \sim \frac{K_2(0)}{1- \omega^2/\omega_f^2}
\end{align}
where $\omega_f/(2\pi)$ is equal to the $f$-mode frequency of the star. 
This frequency is typically greater than contact frequency, and it lies at or above $1.8 
\,\mathrm{kHz}$.
Hence, the orbital frequency of the binary cannot cross the $f$-mode frequency, and the conservative tidal response function continuously increases with increasing frequency.
We observe this behavior clearly in all panels of Fig.~\ref{fig:conservative_tidal_response}.

\begin{figure}[h!]
    \centering
    \includegraphics[width = 0.95 \columnwidth ]{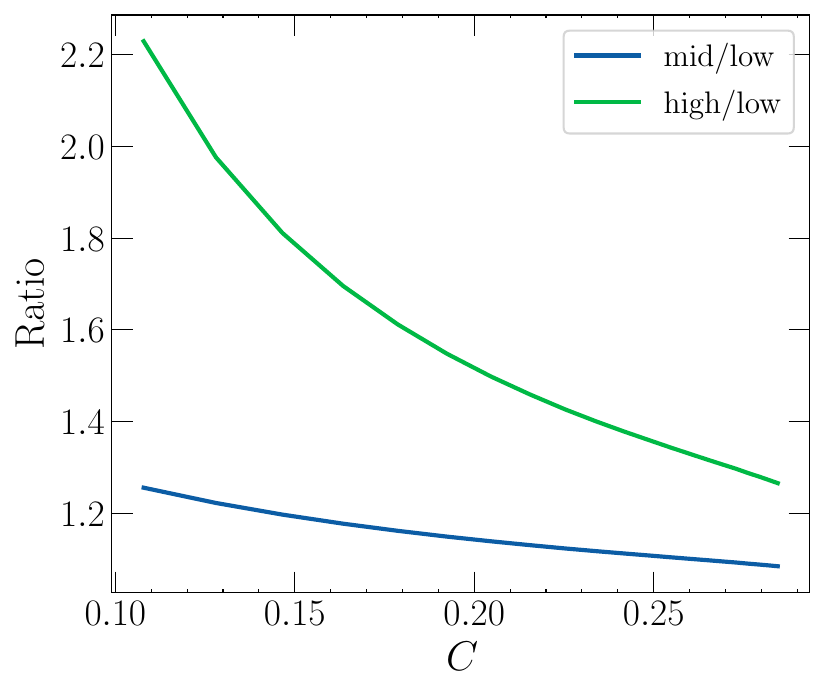}
    \caption{
      Frequency dependence of the conservative tidal response for the $npe$ EoS  as a function of the stellar compactness $C$, with fiducial parameter values $\thetanpe = \thetanpe^{\mathrm{fid}}$ from Eq.~\eqref{eq:theta-canonical}.
      The green line traces the ratio $\mathrm{Re}\left[\hat{K}_2(f_{\mathrm{high}})\right]/\mathrm{Re}\left[\hat{K}_2(f_{\mathrm{low}})\right] $ and the blue line traces the ratio $\mathrm{Re}\left[\hat{K}_2(f_{\mathrm{mid}})\right]/\mathrm{Re}\left[\hat{K}_2(f_{\mathrm{low}})\right] $. Observe that the evolution of the tidal response with frequency is non-negligible, especially for less compact stars. 
    }
    \label{fig:conservative_tidal_response_ratio}
\end{figure}

From Fig.~\ref{fig:conservative_tidal_response}, we see that, on average, $\mathrm{Re}\left[\hat{K}_2(f_{\mathrm{high}})\right] \sim 2 \,\mathrm{Re}\left[\hat{K}_2(f_{\mathrm{low}})\right] $ for stars with $C<0.2$, i.e. the dynamical tidal response can be twice as large as the frequency increases. We quantify this more precisely in Fig.~\ref{fig:conservative_tidal_response_ratio}, where we plot the ratios $\mathrm{Re}\left[\hat{K}_2(f_{\mathrm{high}})\right]/\mathrm{Re}\left[\hat{K}_2(f_{\mathrm{low}})\right] $ for the $npe$ model with the fiducial values $\thetanpe^{\mathrm{fid}}$ as a function of compactness.
Observe that the increase is largest at low compactness, as the stars deform more easily then.
If not accounted for in the waveform, this increase in $K_{2}(\omega)$ can bias the inference of nuclear physics from gravitational-wave observations across a population of neutron stars~\cite{Pratten:2021pro}, especially as higher signal-to-noise ratio events are observed.
This has motivated recent work~\cite{Abac_2024} to incorporate dynamical tidal excitation due to the $f$-mode in waveform models.

Stars with smaller compactness demonstrate the greatest sensitivity to both the orbital frequency and the saturation coefficients. Consequently, the most informative constraints on nuclear matter properties will come from binaries of lighter neutron stars. 
It appears that once a neutron star becomes too dense, predictions from different sets of nuclear parameters become degenerate, as the star does not significantly deform. Of course, one should still investigate this more thoroughly in cases where hyperons or strong phase transitions are present in the stellar core. 

\begin{figure*}[thp!]
    \centering
    \includegraphics[width = 0.95 \columnwidth ]{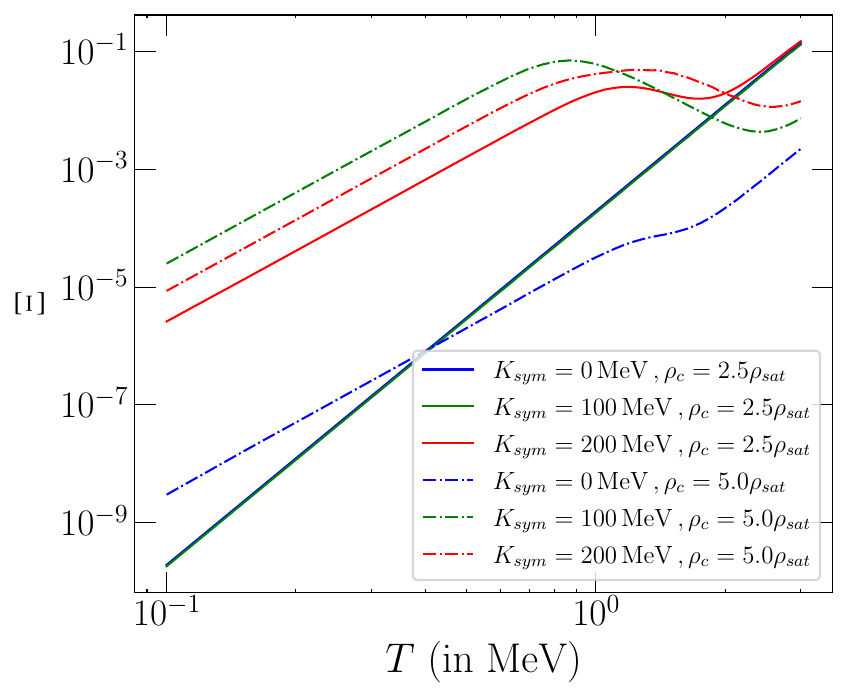}
    \includegraphics[width = 0.95 \columnwidth ]{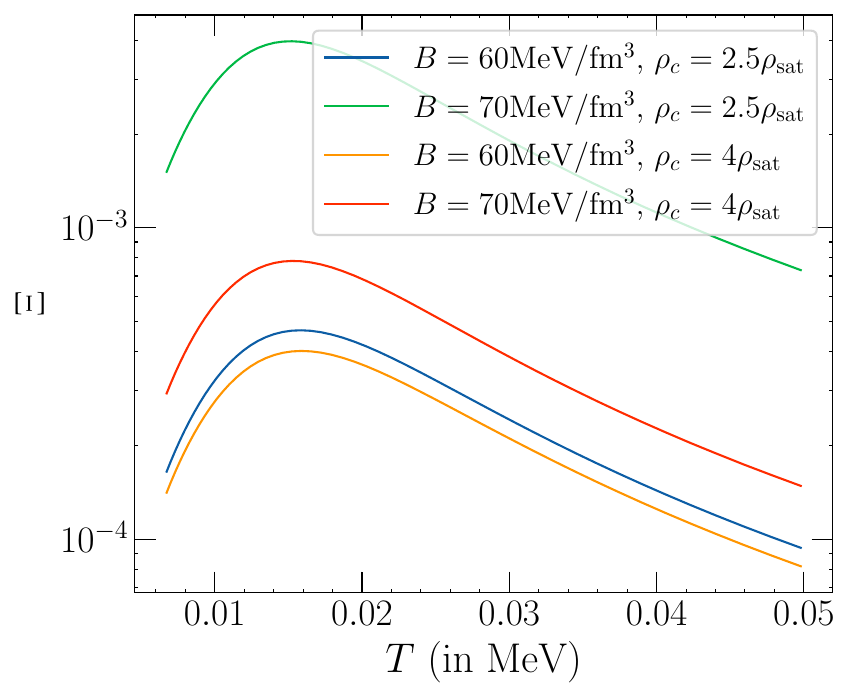}
    \caption{
        Dependence of the dissipative tidal deformability on the temperature of the star.
        The left panel shows the value of the dissipative tidal deformability $\Xi$ at $f=f_{\mathrm{low}}$ within the $npe$ model, for different values of $K_{\mathrm{sym}}$ and of central mass density. The right panel presents the same quantity within the quark-matter model, for different values of the bag constant $B$ and of the central mass density.
    }
    \label{fig:dissipative_response_non_resonant}
\end{figure*}
\subsection{The dissipative tidal response function}\label{sec:dissipative-tidal-response}
We now investigate the dissipative tidal deformability $\Xi$. 
Unlike its conservative counterpart, the dissipative deformability is sensitive to the temperature inside the neutron star, because the bulk viscosity is highly  temperature-dependent.
In the left-panel of Fig.~\ref{fig:dissipative_response_non_resonant} we show the temperature dependence of the dissipative tidal deformability for a few representative values of $K_{\mathrm{sym}}$ and of central density $n_{B,c}$, for a gravitational wave frequency $f = f_{\mathrm{low}}$ and with all other nuclear physics parameters fixed to fiducial values from Eq.~\eqref{eq:theta-canonical}. 
As we see from this figure, the dissipative tidal response shows a steep dependence on temperature and on $K_{\rm sym}$. 

Interestingly enough, relaxation peaks appear at certain temperatures (see the solid red curve at $T\sim 1$ MeV)  in the plots as a function of the temperature.
The origin of these temperature-dependent relaxation peaks can be understood by examining Eq.~\eqref{eq:kubozeta}.
This equation shows that, if the timescale $\tau_\Pi$ for the microscopic process is of the same order as the macroscopic timescale $\frac{1}{\omega}$, the equilibration becomes out-of-phase with the macroscopic perturbations, leading to dissipation. These relaxation peaks in temperature occur when this happens.
The qualitative features observed in the npe EoS are also observed in the right panel, where we present the results for the quark-matter model.

A confident measurement of dissipative tidal deformability for a GW170817-like event in third generation detectors requires $\mathrm{max}\left[\Xi\right] > \mathcal{O}(100)$~\cite{Ripley:2023lsq}. Therefore, the frequency-dependent dissipative tidal deformability shown in Fig.~\ref{fig:dissipative_response_non_resonant} is $1-3$ orders of magnitude smaller than the threshold for measurement with current or future gravitational-wave detectors, suggesting that microphysical sources of viscosity, such as Urca processes in $npe$ matter or strangeness-changing reactions in quark matter cannot lead to measurable imprints on gravitational-wave observables. 
Other physical ingredients such as hyperons or strangeness dominated phases \cite{Cruz-Camacho:2024odu}, color superconducting phases, or the presence of phase transitions may lead to different conclusions, but their investigation is left to future studies. 
There are also some suggestions that shear viscosity could lead to measurable imprints on the gravitational-wave signals~\cite{Shterenberg:2024tmo,Saketh:2024juq}.
However, we stress that non-microscopic sources such as effective turbulence may also contribute to the dissipative tidal response function.
\subsection{Low-frequency $g$-mode resonances}\label{sec:g-mode-resonances}
\begin{figure*}[thp!]
    \centering
    \includegraphics[width = 0.95 \columnwidth ]{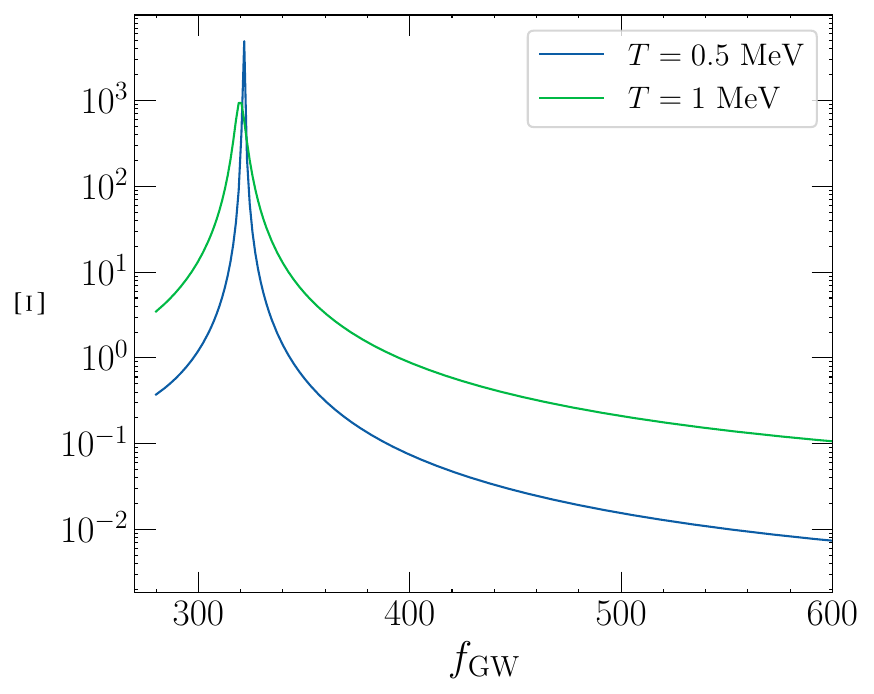}
    \includegraphics[width = 0.95 \columnwidth ]{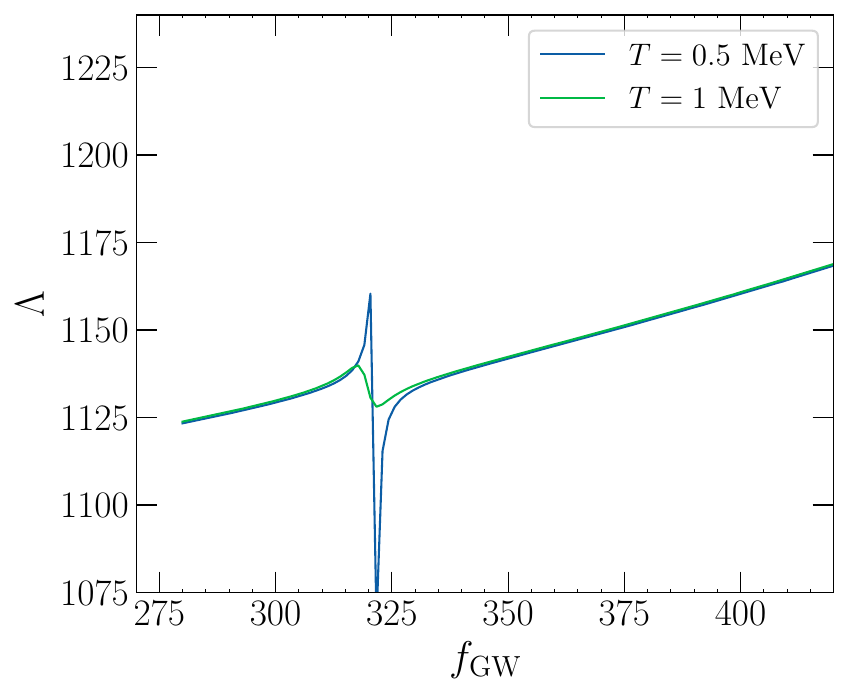}
    \caption{
    Low-frequency resonances in the tidal response function for the $npe$ EoS with parameters $\left(\rho_{c}, K_0, S_{\mathrm{sym}}, L_{\mathrm{sym}}, K_{\mathrm{sym}} \right) = (2.2 \rho_{\mathrm{sat}}, 250 \mathrm{MeV}, 32 \mathrm{MeV}, 120 \mathrm{MeV}, -50 \mathrm{MeV} )$ for two different values of temperature.
    The left-panel shows the peak in the dissipative tidal response function due to the $g$-mode resonance at $320 \mathrm{Hz}$.
    The right panel shows the change in the conservative tidal response function due to the same $g$-mode resonance.
    Most $g$-mode resonances are in the range $f_{\mathrm{GW}}<300 \mathrm{Hz}$, but here the strong stratification caused due to PREX-II-like values for the nuclear parameters pushes the $g$-mode resonances to frequencies higher than $300 \mathrm{Hz}$.
    }
    \label{fig:dissipative_response_g_resonant}
\end{figure*}
The discussion on the tidal response function in the previous subsections has focused on the frequency regime where  $f_{\mathrm{GW}} \geq f_{\mathrm{low}} = 400 \mathrm{Hz}$.
While this restriction is adequate for current detectors, the increased sensitivity of third-generation detectors could allow one to potentially measure low-frequency $g$-mode resonances that arise due to stratification inside neutron stars~\cite{Kokkotas:1999bd}.
These $g$-mode resonances have been modeled extensively using Newtonian models of the tidal response function (see eg.~\cite{Lai:1993di,Yu_2016,Kuan_2021,Yu:2022fzw,Counsell_2024,Gao:2025aqo} and references therein).
One generally expects them to cause a dephasing in the gravitational-wave signal of  $\mathcal{O}\left(10^{-3} \mathrm{rads}\right)$, thus making them potentially detectable by very high signal-to-noise ratio events in third-generation detectors~\cite{Yu_2016,Counsell_2024}. We note that at higher temperatures the weak-interaction relaxation rate becomes comparable to the $g$-mode frequencies, so that the mode frequencies (with quasi-normal mode boundary conditions) become noticeably complex: the real part decreases and the imaginary part grows, and the $g$-modes are eventually suppressed entirely by bulk-viscous damping~\cite{Zhao:2025pgx}. 
For the temperatures considered in this work ($T \lesssim 1\,\mathrm{MeV}$ for most of our parameter scan, with $T$ reaching up to a few MeV in the $npe$ case), this suppression is mild and our treatment of $g$-modes as well-defined resonances of the tidal response remains appropriate; the overall trend of weakening $g$-mode resonance with increasing $T$ is consistent with the temperature behavior we observe.

To understand $g$-mode resonances, it is beneficial to use a simple model of the tidal response  from Newtonian theory.
Suppose that the dominant resonances are the first-order $g$-mode and $f$-mode ones. Then, we can model the tidal response function as~\cite{HegadeKR:2025qwj}
\begin{align}\label{eq:g-mode-resonance-ansatz}
    \hat{K}_2(\omega) \sim \frac{k_{2,\mathrm{g-mode}}}{1 - i \tau_{\mathrm{g}} \omega - \omega^2/\omega^2_{\mathrm{g}}} + 
    \frac{k_{2,\mathrm{f-mode}}}{1 - i \tau_{\mathrm{f}} \omega - \omega^2/\omega^2_{\mathrm{f}}}
\end{align}
where the constants $k_{2,\mathrm{g-mode}}$ and $k_{2,\mathrm{f-mode}}$ are the contribution to the adiabatic tidal response function due to $g$- and $f$-mode resonances, $\omega_{g,f}$ are the mode frequencies, and $\tau_{g,f}$ are damping coefficients, with the dimensionless product $\tau_{g,f} \omega_{g,f}$ setting the strength of the damping (equivalently the inverse quality factor, $Q_{g,f}^{-1} = \tau_{g,f} \omega_{g,f})$.
Typically, we have that $\omega_f \gg \omega_g$, $k_{2,\mathrm{g-mode}} \ll k_{2,\mathrm{f-mode}}$, and $(\tau_{\mathrm{f}} \omega,\tau_{\mathrm{g}} \omega) \ll 1$.

Let $\epsilon = {\cal{O}}(\tau_{\rm f,g} \omega)$ be a small positive constant, near the $g$-mode resonance. Then, we can approximate the tidal response function in Eq.~\eqref{eq:g-mode-resonance-ansatz} as
\begin{align}
    \hat{K}_2(\omega_{g} \left(1 \pm \epsilon\right)) 
    &
    \approx
    i \frac{k_{2,\mathrm{g-mode}}}{\tau_{\mathrm{g}} \omega_g} 
    \nonumber\\
    &\mp
    \frac{2 k_{2,\mathrm{g-mode}} \epsilon}{\tau_g^2 \omega_g^2}
    +
    k_{2,\mathrm{f-mode}}
    \,.
\end{align}
From the above equation, we see that, even though $\tau_{\mathrm{g}} \omega_g $ may be a small number, the imaginary piece corresponding to the dissipative tidal deformability near the $g$-mode resonance is enhanced due to its damping,
\begin{align}\label{eq:near-g-mode-resonance}
    \left|\mathrm{Im} \left[ \hat{K}_2(\omega_{g}) \right]
    \right|
    \approx
    \frac{k_{2,\mathrm{g-mode}}}{\tau_{\mathrm{g}} \omega_g} 
    \gg k_{2,\mathrm{g-mode}}\,.
\end{align}
Typical values of the $g$-mode frequency are in the range $f_{\mathrm{g-mode}}  < 300 \mathrm{Hz}$ and resolving this regime with our present code is challenging.
However, the value of the $g$-mode frequency increases with the amount of stratification inside the star. 
Toward the stiff, large-$L_{\mathrm{sym}}$ end of the allowed range (which we illustrate using PREX-II-like values), stratification increases significantly, and shifts the value of the $g$-mode frequency to higher values as we discuss below. This behavior is physically expected: the buoyancy supporting these composition $g$-modes arises from the gradient in the equilibrium composition through the star, a well-established mechanism for neutron-star matter~\cite{1992ApJ...395..240R}, and is the reactive counterpart of the weak-interaction bulk viscosity, which for $npe$ matter is set in the $T\to0$ limit by the density dependence of the symmetry energy~\cite{Yang:2025yoo}. The strong sensitivity of the $g$-mode frequency to $L_{\rm sym}$ is therefore a robust trend, whereas the specific value ($\sim 320$ Hz) is illustrative, since the $g$-mode is a global oscillation whose precise frequency depends on the full stellar background and the assumed particle content; effects such as the muonic channel we neglect would shift it while leaving the qualitative $L_{\rm sym}$ trend intact.

Figure~\ref{fig:dissipative_response_g_resonant} shows the tidal response function for $\left(\rho_{c}, K_0, S_{\mathrm{sym}}, L_{\mathrm{sym}}, K_{\mathrm{sym}} \right) = (2.2\, \rho_{\mathrm{sat}}, 250 \,\mathrm{MeV}, 32 \,\mathrm{MeV}, 120 \,\mathrm{MeV}, -50\, \mathrm{MeV} )$ and for two different temperatures.
The left and right panels show the dissipative and conservative tidal response functions, respectively.
Note that the dissipative tidal response peaks near the $g$-mode resonance at $f_{\mathrm{g}}\sim 320 \mathrm{Hz}$. 
As can be inferred from Eq.~\eqref{eq:g-mode-resonance-ansatz}, a more pronounced peak results from a weaker dissipation (smaller $\tau_g\,\omega_g$), which simultaneously implies smaller dissipative deformability away from the resonance. 
Observe also that the resonance impacts the conservative tidal response function too, as shown in the right panel.
\section{Conclusions}\label{sec:conclusions}
In this work, we used the relativistic tidal-response framework of Refs.~\cite {Poisson:2020vap,Pitre:2023xsr,HegadeKR:2024agt} to compute, in a single consistent setup, the frequency-dependent \emph{conservative} and \emph{dissipative} tidal response of neutron stars from microphysical input. On the microscopic side, we introduced simple analytic models: for neutrino-transparent $npe$ matter, we constructed a meta-model parameterized by the symmetry-energy coefficients $(S_{\rm sym},L_{\rm sym},K_{\rm sym})$, while for quark matter, we employed the MIT bag model. On the macroscopic side, we extended earlier Newtonian studies~\cite{Lai:1993di,Yu_2016} by treating the full complex tidal response without invoking a low-frequency expansion, thereby enabling a direct and transparent connection between nuclear parameters, transport physics, and the dynamical tidal imprint relevant for late inspiral.

Our main results answer the three questions posed in the Introduction and sharpen the physical message. First, we found that the conservative response is controlled by an effective combination of symmetry-energy parameters for $npe$ matter. 
While changes in $L_{\rm sym}$ lead to the largest sensitivity in the conservative response, its range is more tightly constrained due to experimental results as compared to $K_{\rm sym}$. Thus,  the largest absolute variation occurs
along the $K_{\rm sym}$ direction for our fiducial parameter set,
primarily because $K_{\rm sym}$ remains largely unconstrained and
can be scanned over a wide range. However, for the stiff, large‑$L_{\mathrm{sym}}$ (PREX‑II‑like) EoS, $L_{\rm sym}$ produces comparably significant
shifts in the tidal response even over a narrower range of
variation. The relative importance of each coefficient, therefore,
depends on the region of parameter space under consideration
and on the allowed range of variation, underscoring the need
for a joint analysis of symmetry-energy parameters rather than
a one-at-a-time ranking. 
This is especially notable because $K_{\rm sym}$ remains largely unconstrained by experiment: the dynamical tidal response therefore carries information on this sector of the symmetry energy that is accessible to gravitational-wave observations. Establishing the extent to which the symmetry-energy coefficients can actually be inferred requires a dedicated parameter-estimation study, which we leave to future work.

We also showed that the composition $g$-mode frequencies are sensitive to $L_{\rm sym}$, increasing toward the stiff, large-$L_{\rm sym}$ end of the currently allowed range. This is physically expected, since the $g$-mode buoyancy is set by the equilibrium composition gradient~\cite{1992ApJ...395..240R}---the reactive counterpart of the bulk viscosity, which in $npe$ matter is controlled in the $T\to0$ limit by the symmetry energy~\cite{Yang:2025yoo}---and the specific values we quote (of order $320$~Hz for PREX-II--like parameters) are illustrative, since the $g$-mode is a global mode whose frequency also depends on the full stellar background and the assumed particle content.
For the quark-matter model, the conservative tidal response shows strong variation with the bag constant (see the bottom panel of Fig.~\ref{fig:conservative_tidal_response}), with the largest variation occurring during the late inspiral. 

Second, we included the dissipative tide both to present the full complex tidal response within a single relativistic framework and to test whether weak-interaction-driven bulk viscosity leaves an observable imprint; we demonstrated how microphysical bulk-viscous processes map into a dissipative tidal response.
By incorporating Urca processes in $npe$ matter and strangeness-changing reactions in quark matter through a composition-evolution equation, and by relating these rates to a frequency-dependent bulk viscosity using a retarded Green's function, we showed that the dissipative tidal deformability is strongly temperature dependent.

Third, we quantified whether weak-interaction--driven bulk viscosity can generate an observable gravitational-wave signature through tidal dissipation. For the EoS and temperatures explored here, we find that the resulting dissipative tidal deformability is several orders of magnitude too small to affect current observations and remains below the level expected to be detectable, even with third-generation instruments, including in the presence of temperature-dependent relaxation-peak structure. 
This null result is itself informative: if future data were to indicate appreciable tidal dissipation, it would point to additional microphysical channels (e.g., extra particle species like hyperons, superfluid degrees of freedom, shear-viscous or mutual-friction mechanisms) or to effective macroscopic dissipation such as turbulence, rather than to the weak-interaction bulk-viscosity channels considered here.

Finally, our analysis highlights a key open theoretical point for making tidal dissipation as predictive as possible. The imaginary part of the microscopic retarded Green's function is directly tied to bulk viscosity. However, the precise and general mapping between the microscopic Green's function used in transport calculations and the macroscopic tidal response function remains unclear. The strong dependence on the EoS we observe suggests that the real (reactive) part of the microscopic response already encodes much of the structure seen in the conservative dynamical tidal deformability. This hints at a deeper correspondence between microscopic and macroscopic response functions.

In conclusion, in this paper, we have established a relativistic framework in which the frequency-dependent conservative and dissipative tidal response can be computed consistently from a specified microphysical model. This framework can be extended systematically. On the microphysics side, one can incorporate more realistic EoSs, additional degrees of freedom (e.g., hyperons, hybrid stars, color superconducting phases, etc.), superfluidity, and more complete treatments of far-from-equilibrium transport. On the macrophysics side, the formalism can be generalized to include crust elasticity~\cite{Gao:2025aqo}, phase transitions~\cite{Counsell:2025hcv}, stellar rotation~\cite{Yu:2024uxt}, and non-linear tidal interactions~\cite{Yu:2022fzw,Kwon:2024zyg,Kwon:2025zbc,Pani:2025qxs}. 

A particularly interesting avenue for future study is understanding the impact of $i$-modes on dynamical tides. Several recent studies have proposed that measurement of $i$-modes across a population of neutron stars~\cite{Counsell:2025hcv,pereira2025dynamicaltidesneutronstars,Gao:2025aqo} could help in constraining first-order phase transitions inside neutron stars. Combining this dynamical tidal information from gravitational waves with X-ray mass--radius measurements could provide a particularly powerful multi-messenger route for understanding quark-hadron phase transitions inside neutron stars. The dynamical tidal response framework developed here provides a relativistic approach to study the impact of $i$-modes on the gravitational waves emitted by binary neutron star systems.
We leave these important extensions to future study. 

\section*{Acknowledgments} 
We thank E.~Speranza for discussions about bulk viscosity of npe matter, M. Alford, L. Gavassino, S. Harris, C. Manuel, E. R. Most, and L. Tolos for their detailed comments on an earlier version of the draft. Many of the authors here are supported by the NSF under the MUSES collaboration
OAC2103680. MH was supported by the Brazilian National Council for Scientific and Technological Development (CNPq) under process No. 313638/2025-0.
JNH, JN, and YY acknowledge the US-DOE Nuclear
Science Grant No. DE-SC0023861.
NY acknowledges support from the Simons Foundation through
Award No. 896696, the Simons Foundation International
through Award No. SFI-MPS-BH-00012593-01, the NSF
through Grants No. PHY-2207650 and PHY-25-12423,
and NASA through Grant No. 80NSSC22K0806.
\appendix 
\input{appendix}

\bibliography{ref}
\end{document}

%% file: appendix.tex
\section{Calculation of thermodynamic derivatives}\label{appendix:derivatives}
Let $J$ be an arbitrary function of $(T, \rho,Y_X)$.
We can obtain derivatives with respect constant $s$, such as, $\left. \frac{\partial J }{\partial \rho} \right|_{s, Y_X}$ as follows.
We start by differentiating $ds$, which is zero for isentropic perturbations considered in the main text 
\begin{align}
    0 = ds = \left. \frac{\partial s }{\partial T} \right|_{\rho, Y_X}
    d T
    +
    \left. \frac{\partial s }{\partial \rho} \right|_{T, Y_X}
    d \rho 
    +
    \left. \frac{\partial s }{\partial Y_X} \right|_{\rho, Y_X}
    d Y_X
    \,,
\end{align}
to obtain
\begin{align}\label{eq:dT-when-ds-eq-0}
    d T
    &=
    -
    \left[ 
    \left. \frac{\partial s }{\partial T} \right|_{\rho, Y_X}
    \right]^{-1}
    \left\{
    \left. \frac{\partial s }{\partial \rho} \right|_{T, Y_X}
    d \rho
    +
    \left. \frac{\partial s }{\partial Y_X} \right|_{T, \rho}
    d Y_X
    \right\} \,.
\end{align}
We now obtain the differential of $dJ$ 
\begin{align}
    dJ
    &=
    \left. \frac{\partial J }{\partial T} \right|_{\rho, Y_X}
    d T
    +
    \left. \frac{\partial J }{\partial \rho} \right|_{T, Y_X}
    d \rho 
    +
    \left. \frac{\partial J }{\partial Y_X} \right|_{\rho, T}
    d Y_X \,, 
\end{align}
and use Eq.~\eqref{eq:dT-when-ds-eq-0} to get
\begin{align}
    \left. \frac{\partial J }{\partial \rho} \right|_{s, Y_X}
    &=
    \left. \frac{\partial J }{\partial \rho} \right|_{T, Y_X}
    -
    \left[ 
    \left. \frac{\partial s }{\partial T} \right|_{\rho, Y_X}
    \right]^{-1}
    \left. \frac{\partial J }{\partial T} \right|_{\rho, Y_X}
    \left. \frac{\partial s }{\partial \rho} \right|_{T, Y_X} \,,\\
    \left. \frac{\partial J }{\partial Y_X} \right|_{s, \rho}
    &=
    \left. \frac{\partial J }{\partial Y_X} \right|_{T, \rho}
    -
    \left[ 
    \left. \frac{\partial s }{\partial T} \right|_{\rho, Y_X}
    \right]^{-1}
    \left. \frac{\partial J }{\partial T} \right|_{\rho, Y_X}
    \left. \frac{\partial s }{\partial Y_X} \right|_{T, \rho}
    \,.
\end{align}
\section{Urca rates}\label{appendix:Urca}
For the $npe$ matter, we use the Fermi surface approximation and obtain analytical expressions for the Urca rates \cite{Alford:2018lhf, Alford:2021ogv, Harris:2020rus} as a sum of the direct Urca and the modified Urca rates, 
\begin{equation}
    \Ga_e = \Ga_{dU} + \Ga_{mU,n} + \Ga_{mU,p},
\end{equation}
where
\begin{widetext}
\begin{subequations}
    \begin{align}
        &\Ga_{dU} = \frac{G^2(1+3g_A^2)}{240\pi^5}E^*_{Fn}E^*_{Fp}p_{Fe}\vartheta_{dU}\de\mu_Q(17 \pi^4T^4 + 10\pi^2\de\mu_Q^2T^2 + \de\mu_Q^4) , \\
        &\Ga_{mU,n} = \frac{1}{5760\pi^9}G^2 g_A^2 f^4 \frac{(E^*_{Fn})^3 E^*_{Fp}}{m_\pi^4} \frac{p_{Fn}^4 p_{Fp}}{(p_{Fn}^2+m_\pi^2)^2}\vartheta_n \de\mu_Q( 1835 \pi^6 T^6 + 945\pi^4 \de\mu_Q^2 T^4 + 105\pi^2\de\mu_Q^4 T^2 + 3\de\mu_Q^6 ) ,\\
        &\Ga_{mU,p} = \frac{1}{40320\pi^9}G^2 g_A^2 f^4 \frac{E^*_{Fn} (E^*_{Fp})^3}{m_\pi^4} \frac{p_{Fn}(p_{Fn}-p_{Fp})^4}{((p_{Fn}-p_{Fp})^2+m_\pi^2)^2}\vartheta_p \de\mu_Q \bigg( 1835\pi^6 T^6 + 945\pi^4 \de\mu_Q^2 T^4 + 105\pi^2\de\mu_Q^4T^2 \nonumber\\
        &\hspace{12cm}+ 3\de\mu_Q^6 \bigg),
    \end{align}
\end{subequations}
\end{widetext}
where the pion-nucleon coupling constant $f\approx1$, $G^2 = G_F^2 \,cos^2\theta_c = 1.1\times 10^{-22} $MeV$^{-4}$, $G_F$ is the Fermi coupling constant, $\theta_c$ is the Cabibbo angle, the axial vector coupling constant is $g_A = 1.26$, $p_{FN}$ is the nucleon Fermi momentum, and $E^*_{FN} = \sqrt{p_{FN}^2 + m_N^{*2}}$ is the nucleon energy. 

Within this approximation, the direct Urca reaction is only turned on if the following criteria are satisfied,
\begin{equation}
    \vartheta_{dU} =
  \begin{cases}
  1 & \text{if $p_{Fn}<p_{Fp} + p_{Fe}$} \\
  0 & \text{if $p_{Fn}>p_{Fp} + p_{Fe}$}.
  \end{cases}
\end{equation}
The modified Urca rate is always turned on but modified by the density with a factor,
\begin{subequations}
    \begin{align}
        \vartheta_n &=
        \begin{cases}
        1 \quad \text{if $p_{Fn}>p_{Fp} + p_{Fe}$} \\
        1 - \frac{3}{8} \frac{(p_{Fp} + p_{Fe} - p_{Fn})^2}{p_{Fp}p_{Fe}} & \text{if $p_{Fn}<p_{Fp} + p_{Fe}$} ,
        \end{cases} \\
        \vartheta_p &=
        \begin{cases}
        0 \quad \text{if $p_{Fn}>3p_{Fp} + p_{Fe}$} \\
        \frac{(3p_{Fp} + p_{Fe} - p_{Fn})^2}{p_{Fn}p_{Fe}} \quad \text{if} 
        \begin{cases}
        p_{Fn}>3p_{Fp} - p_{Fe} \\
        p_{Fn}<3p_{Fp} + p_{Fe}
        \end{cases} \\
        4 \frac{3p_{Fp} - p_{Fn}}{p_{Fn}} 
        \quad
        \text{if}
        \begin{cases}
        3p_{Fp} - p_{Fe} > p_{Fn} \\
        p_{Fn} > p_{Fp} + p_{Fe}
        \end{cases} \\
        2 + 3\frac{2p_{Fp} - p_{Fn}}{p_{Fe}} - 3\frac{(p_{Fp} - p_{Fe})^2}{p_{Fn}p_{Fe}}  \text{if $p_{Fn}<p_{Fp} + p_{Fe}$} .
        \end{cases}
    \end{align}
\end{subequations}
\section{Strangeness-changing rates}\label{appendix:quark_rates}
\begin{widetext}
For quarks, we only consider the dominant reaction, the strangeness-changing reaction defined by,
\begin{equation}
    u + d \longleftrightarrow u + s.
\end{equation}

If one assumes $T = m_s = 0$, $\mu_d = \mu_u$, and $\nu \equiv \frac{\mu_s - \mu_d}{\mu_u} \to 0$, a simple analytical expression can be obtained \cite{Madsen:1993xx},
\begin{equation}
    \Ga(s u_1 \to u_2 d) \approx \frac{16}{5\pi^2} G_F^2 \sin^2\theta_C \cos^2 \theta_C \mu_u^5 (\mu_s - \mu_d)^3 .
\end{equation}

The finite temperature and finite mass effects can be accounted for through a correction term \cite{Madsen:1993xx},
\begin{equation}
    \frac{16}{5\pi^2} G_F^2 \sin^2\theta_C \cos^2 \theta_C \mu_u^5 (\mu_s - \mu_d) 4\pi^2 T^2 \chi,
\end{equation}
where $z \equiv \mu_d/\mu_u$, and 
\begin{equation}
    \chi =
    \begin{cases}
    -\frac{1}{128}\nu^5 - \frac{5}{64}z\nu^4 + \frac{5}{64} (2-z^2)\nu^3 + \frac{5}{16}(1+3z)\nu^2 + \frac{5}{16}(4z + 3z^2)\nu + \frac{5}{4}z^2 - \frac{1}{4}  \quad \text{at }m_s=0, \\
    \frac{5}{256} \left( \frac{3}{5}(32 + y^5) - (8 + y^3)\{ 4 + \left[ (\mu_s + \mu_d)/\mu_u \right]^2 - m_s^2/\mu_u^2 \} + (24 + 12y)\{ \left[ (\mu_s + \mu_d)/\mu_u \right]^2 - m_s^2/\mu_u^2 \}\right)  \,\text{at }m_s \neq 0, 
\end{cases}
\end{equation}
\end{widetext}
where 
\begin{equation}
    y = \frac{(\mu_s^2 - m_s^2)^{1/2} - \mu_d}{\mu_u} . 
\end{equation}